\documentclass[pra,aps,twocolumn,10pt,notitlepage,longbibliography]{revtex4-2}
\usepackage{graphicx}
\usepackage{physics}
\usepackage[capitalise]{cleveref}
\usepackage{tikz}
\usetikzlibrary{patterns,arrows,calc,decorations.pathmorphing, shapes.geometric,backgrounds}

\begin{document}
% \title{Complete state description of optomechanical cooling and entanglement from the Lindblad master equation}
\title{Optimization of optomechanical cooling and entanglement using semi-analytic solutions to the Lindblad master equation}
%\title{Optimizing optomechanical entanglement using a complete state description}
\author{Paul R. B. Hughes}
\email{p.hughes@queensu.ca}
\author{Marc M. Dignam}
\affiliation{Department of Physics, Engineering Physics and Astronomy,
Queen's University, Kingston, ON, K7L 3N6, Canada}

\begin{abstract}
We solve the Lindblad master equation for the quantum state of a pumped optomechanical system coupled to a thermal bath. 
%that interacts with a strong microwave pump field. 
We show that when the microwave pump field frequency is on the red sideband of the cavity resonance, the exact form of the state is a beam-split thermal state, and when it is on the blue sideband, it is a two-mode squeezed thermal state. These solutions allow us to determine the mechanical cooling and the entanglement between the microwave and mechanical modes. %in the red and blue sidebands. 
We find that we can entangle the modes in a thermal environment by activating the two fields sequentially. In this scheme, after cooling the mechanical mode via the red sideband pump, we determine the optimal blue sideband pump field to achieve the maximum entanglement time and the maximum time below a desired correlation variance threshold as a function of the loss rates and equilibrium temperatures of the two modes. 

\end{abstract}

\maketitle

\section{Introduction}
Quantum interaction between optical or radio-frequency cavities and micro or nanomechanical resonators has become a promising area in the development of nonclassical states as a resource for quantum metrology and computing \cite{Zhang:24,mechOsc}.
Previous theoretical research has studied the use of an optomechanical interaction to cool mechanical oscillators to the ground state\cite{PhysRevLett.101.263602,PhysRevA.85.051803,PhysRevLett.99.093902}, generate highly squeezed mechanical modes \cite{PhysRevA.88.063833,https://doi.org/10.1002/qute.201800110,PhysRevA.110.013512} and entangle microwave cavities to mechanical motion \cite{PhysRevLett.110.253601,PhysRevA.84.052327}.
Several experiments have verified the cooling \cite{teufelCooling,electromechCooling}, %coherent state population transfer \cite{coherentStateTransfer}, 
entanglement \cite{doi:10.1126/science.1244563}, and squeezing \cite{mechOsc,kerrMech} properties of the interaction, allowing researchers to measure mechanical motion to a higher precision than quantum shot noise allows \cite{PhysRevLett.123.183603}. %In a similar setup, membrane-in-the-middle schemes using optomechanical interactions have attracted recent attention for entangling two optical modes in integrated systems \cite{PhysRevA.86.033821,PhysRevA.87.033829,AbutalebiB.A.:24,Jayich_2008,Agrawal:24}. 

In the systems discussed above, the optomechanical interactions are generated using a nonlinear response to the pumping of the system with a coherent electromagnetic field operating on one of the mechanical sidebands of the cavity resonance. When the cavity field is on the red sideband, the Hamiltonian takes the form of a beam-splitter interaction, %cite
while when it is on the blue sideband, it takes the form of a spontaneous parametric down-conversion (SPDC) process, such as that seen in an optical parametric oscillator. In both cases, the system is also interacting with a thermal bath, such that the equilibrium thermal occupancy of the system modes must be considered.
Previous authors have theoretically studied the entanglement of two-mode squeezed thermal states coupled to environments \cite{entanglementTMSTS,TMSVSres,PhysRevA.78.052313,PhysRevLett.100.220401}, solving the free evolution of the system coupled to a thermal or squeezed bath. 
%Due to the importance of the environment to the optomechanical system, a reservoir engineering approach is frequently used to generate the desired system state \cite{PhysRevB.70.205304,PhysRevA.89.063805}. %In these systems, the thermal environment and its coupling to the system is manipulated to generate quantum resources.

Cavity optomechanics is typically analyzed by starting with the optomechanical Hamiltonian and the Langevin equations for stochastic loss to determine the steady state characteristics of the systems. Input-output formalism is used, where loss channels, laser pumping, and the interaction are components of the evolution of the creation operator for each mode \cite{PhysRevLett.101.263602,PhysRevA.85.051803,AYEHU2024100605,PhysRevA.98.063843,PhysRevLett.107.063602}. 
Alternatively, numerical solvers such as QuTiP \cite{JOHANSSON20121760,JOHANSSON20131234} and the Julia Quantum Optics library \cite{kramer2018quantumoptics} can be used to approximate the optomechanical master equation in a limited Fock basis \cite{PhysRevA.109.L051501}. 

In this work, we show that the exact form of the density operator for a strongly pumped optomechanical system is a beam-split thermal state when the pump is on the red sideband and a two-mode squeezed thermal state when the pump is on the blue sideband of the cavity resonant frequency, and we derive the coupled first-order differential equations that govern the evolution of the state parameters. This method offers an advantage over either of the two former methods, as the Langevin equations only follow moments of the creation and annihilation operators, while our exact state description allows for simple calculation of any operator moments from the density operator. %This method considers a time-domain solution, allowing us to analyze the transient system.
%The equations of state presented here can be solved using fewer computational resources than required of the current computational approaches, especially for large reservoir populations coupled to the system.  % only include if directly comparing. Thesis work?

Our method extends the technique first used in Seifoory \textit{et. al.} \cite{hossein}, where the LME for a pumped degenerate optical parametric oscillator was solved, showing that for that system, the state generated is a squeezed thermal state. We employ a similar approach here to model the beam-splitter interaction of red-sideband optomechanics. Our work also extends previous two-mode squeezed thermal state results for an SPDC interaction \cite{PhysRevA.103.022418} to situations where the interaction with a thermal bath is important. %the thermal environment present in the blue-sideband optomechanics. 

We use our approach to model an optomechanical system where the thermal occupation of the microwave modes is negligible before the pump arrives. We find that when the pump is on the red sideband, the maximum cooling of the mechanical mode is largely determined by the loss rate of the mechanical mode. When the pump is on the blue sideband, the loss rate largely determines the minimum correlation variance between the microwave and mechanical modes, which, in turn, determines whether the modes are entangled. Unlike the zero-temperature case of Ref. \cite{PhysRevA.103.022418}, where matched loss rates between the two modes are desirable to maximize the robustness of the entanglement, we find that for a system that has been cooled such that the mechanical population is below the thermal equilibrium population, entanglement is more easily achieved when the mechanical mode %with a larger equilibrium population 
is significantly less lossy than the microwave mode. Finally, we use those analyses to determine the limits of optomechanical entanglement in a thermal environment, and determine the optimal field strength for a desired entanglement requirement.
%We also combine the beamsplitter and SPDC interactions to optimize the cavity field for the longest time under a correlation variance threshold.

The paper is organized as follows. In Sec.~\ref{theory}, we review the optomechanical system and Lindblad Master Equation (LME) for the system evolution. We perform the standard linearization and arrive at the interaction Hamiltonian for pumping on the red and blue sidebands of the cavity. Next, in Sec.~\ref{cooling}, we consider the red-sideband interaction and solve the LME, showing that the state of the optomechanical system when this interaction is applied to a thermal state is a beam-split thermal state (BTS). We use this result to determine the occupancy of the mechanical mode for given pumping strengths and relative loss rates, showing how the mechanical mode is cooled. 
Sec.~\ref{squeezing} contains the derivation of the two-mode squeezed thermal state coupled to the thermal bath for a blue sideband microwave pump. We discuss the effect of the large environmental population on two-mode squeezing. Finally, in Sec.~\ref{scheme} we combine the results of the previous two sections to determine the limits of entanglement for a given system when it has been subject to a cooling-entanglement scheme. %We describe how these limits lead to an optimal pumping strength to create the longest possible entanglement at a given entanglement requirement.
In Sec. VI., we state our conclusions from the analysis and results.

\section{Optomechanical System}  % this might even belong in the introduction section, as its all previous results
\label{theory}
In this section we apply the approximations commonly used for optomechanical systems to derive the Hamiltonian and LME that we will use in the following sections. 
We consider an optomechanical system composed of a mechanical membrane coupled to a microwave cavity. We model the interaction between a resonant mode of the membrane at frequency $\Omega_m$ with a resonant mode of the microwave cavity at frequency $\omega_c$. The unperturbed system Hamiltonian is
\begin{gather}
H = \hbar \omega_c(x) a^\dag a + \hbar \Omega_m b^\dag b,
\end{gather}
where $a^\dag$ ($a$) and $b^\dag$ ($b$) are the creation (annihilation) operators for microwave quanta in the cavity and mechanical quanta in the membrane, respectively, and $x$ is the displacement of the membrane. This displacement modulates the cavity resonant frequency, which can be modelled to first order as $\omega_c(x) \approx \omega_c -\gamma_0 (b + b^\dag)$ where $\omega_c \equiv \omega_c(0)$ and $\gamma_0$ is the real-valued vacuum optomechanical coupling strength. With this coupling term, the Hamiltonian now reads
\begin{gather}
H = \hbar\omega_c a^\dag a + \hbar \Omega_m b^\dag b - \hbar \gamma_0 a^\dag a (b + b^\dag).
\end{gather}
This system interacts with the environment and an external pump driving field that generates a coherent component in the cavity. The coherent drive allows us to linearize the cavity field operator as $a = \bar{\alpha} + d$, where $|\bar\alpha|^2$ is the average photon number in the cavity field. This coherent component is determined by classical equations of motion (not shown) for the input pump that drives the cavity field. Thus, the coherent component only appears in the interaction term of our Hamiltonian \cite{cavityOptomechanicsRMP}. Note that
$\bar\alpha$ includes the contributions of the drive laser, as well as any loss in the mean field, and is assumed to be a general time-dependent, unchirped pulse. In what follows, we will refer to this as the pump field.

We now move into a frame rotating at the frequency $\omega_L$, of the pump laser field by applying the unitary operator $U(t) = \exp(i\omega_L a^\dag a t)$ to transform the Hamiltonian to $H' = UHU^\dag - i\hbar U\partial U^\dag/\partial t$. This gives us
$H' = H_0 + H_I$, where
\begin{gather}
H_0 = -\hbar \Delta d^\dag d + \hbar \Omega_m b^\dag b, \\ H_I = - \hbar \gamma_0 (\bar{\alpha}^* + a^\dag)(\bar{\alpha} + a)(b + b^\dag),  %\\ H_D =  i\hbar\sqrt{\kappa_{in}}(\eta a^\dag + \eta^* a), \nonumber
\end{gather}
where $\Delta \equiv \omega_L - \omega_c$. 
We now expand the interaction part of the Hamiltonian, $H_I$,  in powers of $\bar{\alpha}$. We neglect the term that is independent of $\bar{\alpha}$ because it will be very small relative to the other terms, assuming that the mean field is large. The term that is second order in $\bar{\alpha}$ is given by $|\bar{\alpha}|^2(b + b^\dag)$. We also omit this term, as its only effect is to create a displacement in the mechanical membrane, which can be compensated in the steady state with an appropriate shift in the mechanical displacement coordinate and laser tuning \cite{cavityOptomechanicsRMP}. For a time-dependent $\bar\alpha$, we would need to include some displacement Hamiltonian, although this would not affect the entanglement properties in which we are interested. Thus the only term that we retain is the one that is first order in $\bar{\alpha}$, which is given by 
\begin{gather}
H_I \approx -\hbar \gamma_0(\bar\alpha^* d + \bar\alpha d^\dag)(b + b^\dag).
\end{gather}
This interaction Hamiltonian can be split into two parts:
\begin{gather}
H_I^{\mathrm{red}} = -\hbar \gamma_0 (\bar\alpha^* d b^\dag + \bar\alpha d^\dag b), \label{BSinteraction} \\
H_I^{\mathrm{blue}} = -\hbar \gamma_0 (\bar\alpha d^\dag b^\dag + \bar\alpha^* d b), \label{squeezeint}
\end{gather}
where $H_I^{\mathrm{red}}$ is resonant and dominant when the laser frequency is on the \textit{red sideband} ($\Delta \approx -\Omega_m$) and  $H_I^{\mathrm{blue}}$ is dominant when the laser frequency is on the \textit{blue sideband} ($\Delta \approx \Omega_m$). This leaves us with
\begin{gather}
H' \approx H_0 + H^{\mathrm{red}}_I + H^{\mathrm{blue}}_I,
\label{EqHprime}
\end{gather}
which is the standard starting point for cavity optomechanics and is the Hamiltonian that we will use in what follows.

The microwave and mechanical modes are coupled to a thermal environment at temperature $T_b$, which has a mean photon number of $n_c^b = (\exp(\hbar\omega_c/kT_b) - 1)^{-1}$ and phonon number $n_m^b = (\exp(\hbar\Omega_m/kT_b) - 1)^{-1}$. The Lindblad master equation for the time evolution of the system density operator $\rho(t)$ is given by \cite{OpenQuantum}
\begin{align}
\begin{aligned}
\frac{d}{dt}\rho (t) = -&\frac{i}{\hbar}\comm{H}{\rho} \\
+ &\kappa (n_c^b + 1) D[d](\rho) + \kappa n_c^b D[d^\dag](\rho) \\
+ &\Gamma_m (n_m^b + 1) D[b](\rho) + \Gamma_m n_m^b D[b^\dag](\rho),
\label{master}
\end{aligned}
\end{align}
where $\Gamma_m$ is the power decay constant of the phonons, $\kappa$ the power decay constant of the microwave mode, and 
\begin{equation}
D[F](\rho) \equiv F\rho F^\dag - \frac{1}{2}\acomm{F^\dag F}{\rho} \label{finalHamm}
\end{equation} 
is the dissipator, which accounts for the two-way coupling with the environment. Note that this LME can be used for systems with small optomechanical coupling, though in the ultra-strong coupling regime, alternatives such as the dressed state master equation \cite{PhysRevA.91.013812} or global master equation \cite{PhysRevA.98.052123} should perhaps be used instead.

In the remainder of this work, we assume that the system is sideband resolved such that $4\Omega_m \gg \kappa$. Under this condition, if the laser is tuned onto the red-sideband, then $H_I^{blue}$ will be negligible and if it is tuned onto the blue-sideband then $H_I^{red}$ will be negligible. Thus, when only one pump is active at a time, we can solve the simpler system with a single interaction term. %the pump frequency is either red or blue detuned, such that only $H_I^{\mathrm{red}}$ or $H_I^{\mathrm{blue}}$ contributes to the Hamiltonian. We are therefore assuming that the system is sideband resolved such that $4\Omega_m \gg \kappa$. This condition is usually required to achieve strong cooling, as it has been shown using classical rate balance that the mechanical population is limited by $n_m > \kappa/(4\Omega_m)$. 
In the next two sections, we do just that, showing the cooling effect that results from \cref{BSinteraction} and the optomechanical entanglement that results from \cref{squeezeint}.

\section{Mechanical Cooling and the Beam-splitter Operator}\label{cooling}

We begin by considering the case where the drive laser is on the \textit{red} sideband, with $\Delta \approx -\Omega_m$. Our interaction Hamiltonian takes the form of \cref{BSinteraction}, where we ignore the term $H^{\mathrm{blue}}_I$. This form resembles that of a beam-splitter interaction, which annihilates quanta in one mode and creates quanta in the other, with the interaction being mediated by the pump-field interaction. As has been previously shown, this interaction can transfer a coherent state from the microwave to the mechanical mode \cite{coherentStateTransfer}. %as it transfers population between the microwave and vibrational modes.
In order to see the mechanism of population transfer clearly, we use the general unitary beam-splitter operator \cite{PhysRevA.40.1371,PhysRevA.65.032323}
\begin{gather}
B(\theta, \phi_B) = \exp[i\theta(d^\dag b e^{i\phi_B} + d b^\dag e^{-i\phi_B})],
\end{gather}
where $\theta$ is the beam-splitter mixing angle and $\phi_B$ is the interaction phase.

We propose that due to the beam-splitter interaction Hamiltonian, the laser drive field will evolve an optomechanical system that is initially in thermal equilibrium into a beam-split thermal state, such that the system density operator can be written as
\begin{gather}
\rho(t) = B(\theta(t), \phi_B(t))\rho_T(n_c^{th}(t), n_m^{th}(t))B^\dag(\theta(t), \phi_B(t)),
\label{BSrho}
\end{gather}
where
\begin{align}
\begin{aligned}
\rho_T(n_c^{th}, n_m^{th}) = &\frac{1}{1 + n_c^{th}} \left( \frac{n_{c}^{th}}{1 + n_{c}^{th}} \right)^{d^\dag d} \\
\times &\frac{1}{1 + n_m^{th}} \left( \frac{n_{m}^{th}}{1 + n_{m}^{th}} \right)^{b^\dag b}
\end{aligned}
\end{align}
is a two-mode thermal state, with time-dependent thermal populations $n_{c}^{th}\left(t\right)$ and $n_{m}^{th}\left(t\right)$ for the microwave and mechanical modes.

In Appendix A, we use the Lindbladian of \cref{master} to show that Eq. (\ref{BSrho}) is the \textit{exact solution} to the master equation, as long as the thermal populations, mixing angle and interaction phase obey the coupled differential equations:
\begin{align}
&\frac{dn^{th}_c}{dt} = \kappa\left[n_c^b - n_c^{th}\right]\cos^2\theta + \Gamma_m\left[ n_m^{b} - n_c^{th}\right]\sin^2\theta \label{BSoptpop}, \\
&\frac{dn^{th}_m}{dt} = \kappa\left[n_m^b - n_m^{th}\right]\cos^2\theta + \Gamma_m\left[ n_c^{b} - n_m^{th}\right]\sin^2\theta \label{BSmechpop}, \\
&\begin{aligned}
\frac{d\theta}{dt} = &\frac{\gamma_0}{2}(\bar{\alpha}^* e^{i\phi_B} + \bar{\alpha} e^{-i\phi_B}) \\
&  +\frac{\sin(2\theta)}{2}\kappa \left( \frac{n_c^b}{n_m - n_c} - \frac{n_m + n_c}{2(n_m - n_c)}\right) \\
& -\frac{\sin(2\theta)}{2}\Gamma_m \left( \frac{n_m^b}{n_m - n_c} - \frac{n_m + n_c}{2(n_m - n_c)}\right),
\end{aligned} \label{mixAngle} \\
&\frac{d\phi_B}{dt} = (-\Delta - \Omega_m) - i\gamma_0\frac{\bar{\alpha}^* e^{i\phi_B} - \bar{\alpha} e^{-i\phi_B}}{2\tan(2\theta)}. \label{intPhase}
\end{align}

We now consider a cavity pump field given by $\bar\alpha(t) = \alpha_0(t) e^{i\phi_L}$. The pulse is unchirped, so $\alpha_0(t)$ is real. Furthermore, we define the detuning of the laser field from resonance with red sideband as $\Delta_+ \equiv \Delta + \Omega_m$, and for simplicity, we choose the pump field phase to be $\phi_L = 0$, so that the beam-splitter operator phase $\phi_B$ is relative to the pump field phase. 

Now we introduce the relative loss rate for the two modes as 
\begin{gather}
\zeta \equiv \frac{\Gamma_-}{\Gamma_+}, \label{zeta}
\end{gather}
where $\Gamma_\pm \equiv (\kappa \pm \Gamma_m)/2$. We also introduce the dimensionless time $\tilde{t} = t\Gamma_+$, the average and relative thermal number variables,
\begin{gather}
2\bar{n}_{th} \equiv n_m^{th} + n_c^{th},  2\bar{n}_{b} \equiv n_m^{b} + n_c^{b},\label{bars} \\
\Delta {n}_{th} \equiv n_m^{th} - n_c^{th}, \Delta {n}_{b} \equiv n_m^{b} - n_c^{b}, \label{deltas}
\end{gather}
and the real dimensionless field strength
\begin{gather}
g_r = 2\frac{\gamma_0 \alpha_0}{\Gamma_+}.
\end{gather}
This allows us to rewrite \cref{BSoptpop,BSmechpop,mixAngle} as
\begin{align}
&\begin{aligned}
\frac{dn^{th}_c}{d\tilde{t}} = &(1 + \zeta)n_c^b\cos^2\theta + (1 - \zeta) n_m^{b}\sin^2\theta \\ &- (1 + \zeta\cos(2\theta))n_c^{th}, \label{cavpopdimless} \end{aligned} \\ & \begin{aligned}
\frac{dn^{th}_m}{d\tilde{t}} = &(1 - \zeta)n_m^b\cos^2\theta + (1 + \zeta)n_c^{b}\sin^2\theta \\ &- (1 - \zeta\cos(2\theta))n_m^{th} \label{mechpopdimless}, \end{aligned} \\
&\frac{d\theta}{d\tilde{t}} = \frac{g_r}{2}\cos(\phi_B) + \frac{2\zeta(\bar{n}_b - \bar{n}_{th}) - \Delta n_b}{\Delta n_{th}}\frac{\sin(2\theta)}{2}, \label{thetadimless} \\
&\frac{d\phi_B}{d\tilde{t}} = \frac{g_r}{2} \frac{\sin(\phi_B)}{\tan(2\theta)} - \frac{\Delta_+}{\Gamma_+}. \label{BangleDE}
\end{align}

\begin{figure}[ht]
\includegraphics[width=\columnwidth]{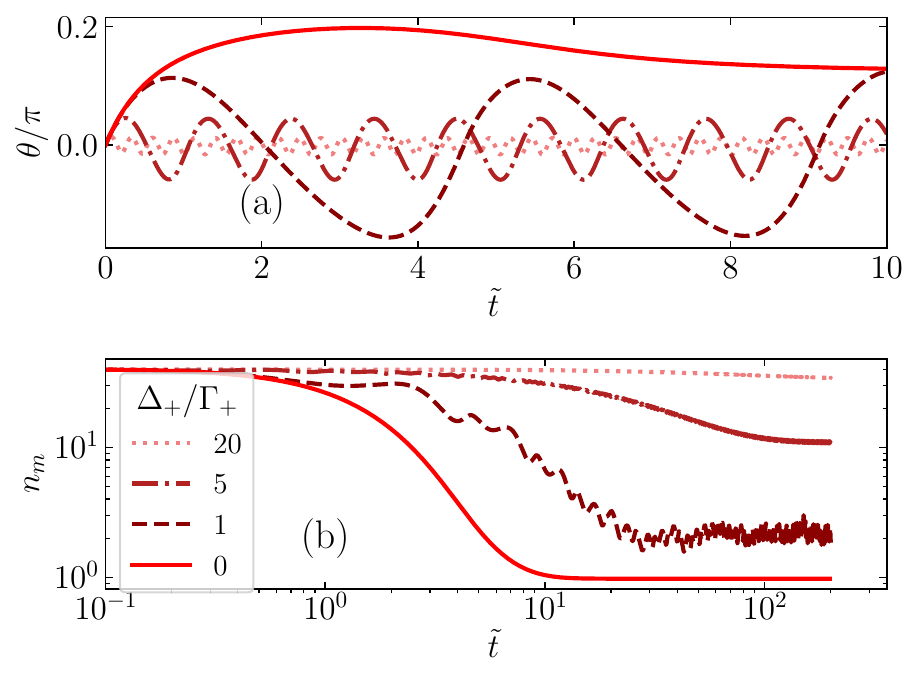}
\caption{The time evolution of (a) the mixing angle $\theta$ and (b) the mechanical mode population for four different values of field detuning from resonance with the red sideband. In all cases, the field strength is moderate ($g_r = 2$) and $\zeta = 0.99$.% and the environmental population is small ($n_m^b = 5, n_c^b = 0$).
}
\label{detuned}
\end{figure}

At this point, we assume that before the pump field is activated at $t=0$, the system is in a two-mode thermal state in the microwave and mechanical modes. %. This assumption is reasonable when the interaction is minimal requiring either $\gamma_0 \ll \Omega_m$, a very small microwave thermal number $n_c^{th}$, or both.
\footnote{The eigenstates for the undriven optomechanical system are exactly given by a two-mode Fock state displaced in the mechanical mode by a strength of $n\gamma_0/\Omega_m$, where $n$ is the photon number \cite{PhysRevA.91.013812}. As long as the microwave bath thermal population is very small or the optomechanical coupling is weak, our assumption that the system starts in a two mode thermal state is excellent.}.
Unless otherwise noted, the plots in this section use $n_m^b = 40$ and  $n_c^b = 0$, corresponding to $\hbar\Omega_m \approx 0.025kT$ and $\hbar \omega_c \gg kT$; for $\Omega_m = 2\pi \times 10$MHz, this corresponds to a temperature of $T\approx 20$mK. 
Because $\theta(0) = 0$, we require $\phi_B(0) = 0$ so that the first term on the right-hand side of \cref{BangleDE} does not diverge at $t=0$. %When there is detuning from the red sideband such that $\Delta_+ \neq 0$, $\phi_B$ evolves from the initial phase of the field. Because $\theta(t)$ will also deviate from zero as time increases, stability requires $\phi_B = \pm n\pi$ at mixing angles $\theta = \pm n\pi/2$.
In \cref{detuned}, we plot the mixing angle $\theta$ and mechanical population as a function of time for a system starting in equilibrium with the environment for various levels of detuning from the red sideband. For the rest of this section, we assume that the pump is activated at $t=0$, and has constant strength $g_r$ for $t>0$.

As can be seen in \cref{detuned}, detuning places limits on the magnitude of $\theta$ and prevents it from reaching a constant value. Thus, detuning slows the cooling ability of the interaction with the field, causing both an oscillation in the mechanical population and a reduction in the cooling. For instance, we can see that a small detuning still allows the mixing angle to rise to the undetuned steady state value, though it fluctuates about this level. Meanwhile, larger detunings prevent $\theta$ from rising to a large value at all, such that one cannot cool down to the single quanta level. This effect is particularly pronounced when the relative loss rate $\zeta$ is not close to 1 or the field is weak. The exact reduction in cooling by detuning can be most easily be quantified by working in the frequency domain \cite{PhysRevLett.99.093902}. In the following, we wish to focus on the cooling \textit{dynamics}, and so we will continue to work in the time-domain. %This will later allow us to examine the decay of the system from the cooled state, though for now we limit our focus to constant fields activated at $\tilde t=0$.
%Our approach will continue to focus on the time-domain for this system.

\begin{figure}[ht]
\includegraphics[width=\columnwidth]{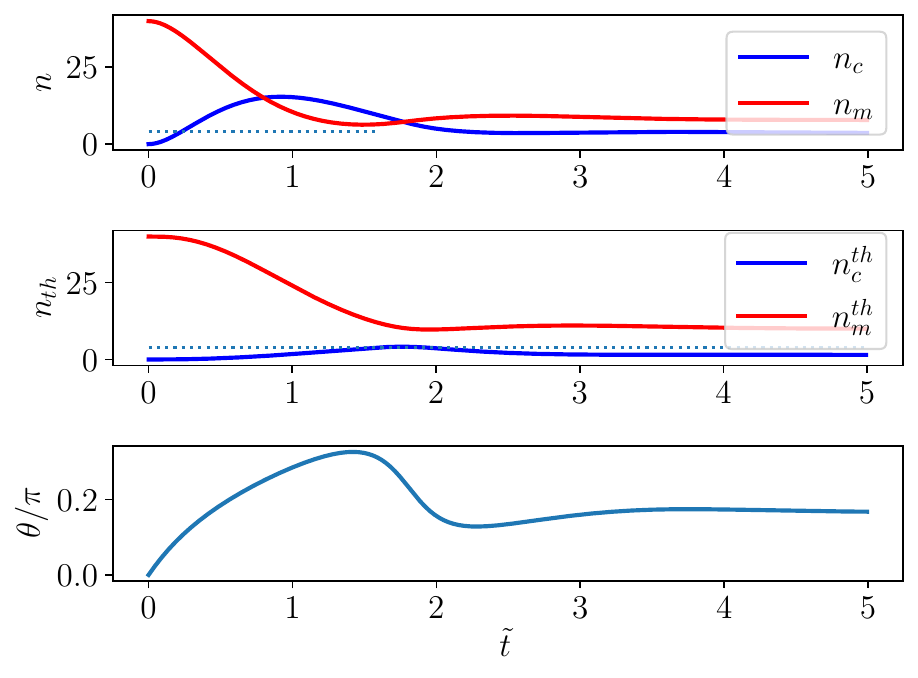}
\caption{Total populations ($n_c$, $n_m$), thermal populations ($n^{th}_c$, $n^{th}_m$), and mixing angle ($\theta$) time evolution during the cooling beam-splitter interaction. The initial state is unmixed ($\theta(0) = 0$) and at equilibrium with the thermal environment. %  with an microwave frequency much greater than the mechanical frequency ($\omega_c \gg \Omega_m$) such that $n_c^b(0) = 0, n_m^b(0) = 5$.
The system has a relative loss ratio $\zeta = 0.8$ and constant pump strength $g_r = 3.5$. The dotted lines indicate the $g_r \rightarrow \infty$ limit.}
\label{temperature}
\end{figure}

In what follows, we consider a system with no detuning ($\Delta_+=0$), so that $\phi_B = \phi_L = 0$ for all time. We investigate the effects of the loss ratio $\zeta$ and the field strength $g_r$ on the system.
As the system evolves, \cref{cavpopdimless,mechpopdimless,thetadimless} force the populations to mix, with the degree of mixing depending on $\zeta$. 
The total population of each mode is a mixture of the thermal populations of the modes. It is easily seen that the populations for the mechanical and microwave modes are given simply by
\begin{gather}
n_m \equiv \expval{b^\dag b} = n_m^{th}\cos^2\theta + n_c^{th}\sin^2\theta, \\
n_c \equiv \expval{a^\dag a} = n_c^{th}\cos^2\theta + n_m^{th}\sin^2\theta.
\end{gather}% in the mechanical system. 
In \cref{temperature}, we plot the total and thermal populations of both the mechanical and microwave modes and the mixing angle of the system as a function of time. 
We see that as the mixing angle grows, the mechanical population is transferred to the microwave mode. However, because the loss rate in the microwave mode is much higher than in the mechanical mode, the additional photons are quickly lost to the environment, and so the microwave population never reaches the initial population in the mechanical mode. When the thermal populations are nearly equal, as is the case just before $\tilde t = 2$, the mixing angle changes rapidly, briefly pushing the populations back towards the thermal equilibrium values, before they settle down to their steady-state values. 

In steady state, the populations force the mixing angle to a stable value, leading the populations to settle to a cooled state in the mechanical mode and an enhanced-population state in the microwave mode. We can determine the steady state populations by setting \cref{cavpopdimless,mechpopdimless,thetadimless} all to zero, which yields the following steady state values for the mixing angle and thermal populations:
\begin{gather}
\tan(2\theta_{ss}) = g_r, \label{steady1} \\
n_c^{th,ss} = \frac{(1 + \zeta)n_c^b \cos^2\theta_{ss} + (1 - \zeta)n_m^b \sin^2\theta_{ss}}{1 + \zeta\cos(2\theta_{ss})}, \label{steady2} \\
n_m^{th,ss} = \frac{(1 - \zeta)n_m^b \cos^2\theta_{ss} + (1 + \zeta)n_c^b \sin^2\theta_{ss}}{1 - \zeta\cos(2\theta_{ss})}. \label{steady3}
\end{gather}

\begin{figure}[ht]
\includegraphics[width=\columnwidth]{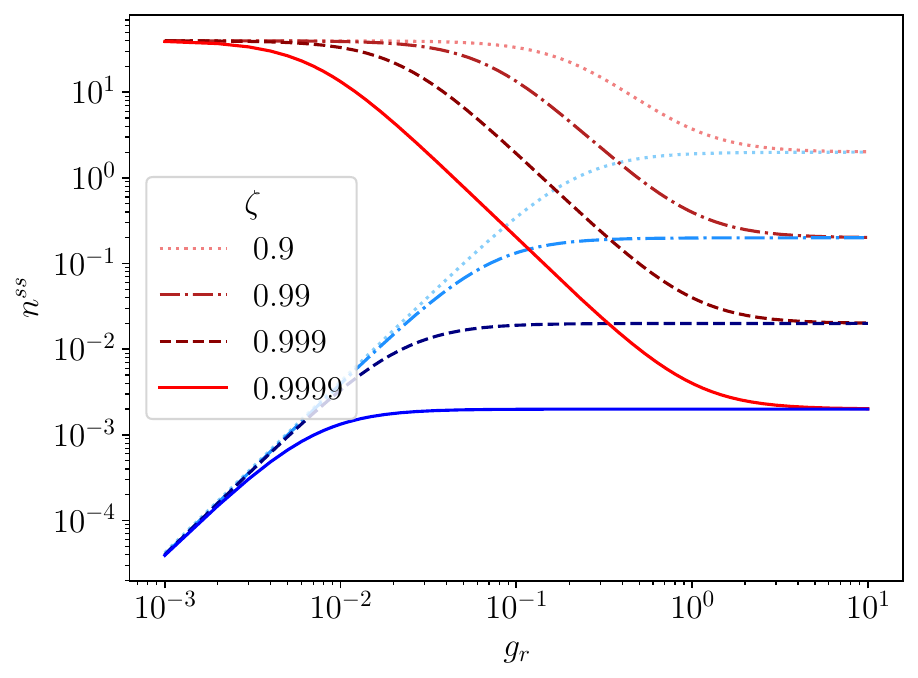}
\caption{Steady state population of the mechanical (red) and microwave (blue) populations, given as a function of the dimensionless pump field strength $g_r$ by \cref{steady1,steady2,steady3}. The four line styles represent different relative loss rates between the microwave and mechanical modes. In the large field limit, the populations converge to a level determined by the loss rate.}
\label{drivevss}
\end{figure}

In \cref{drivevss}, we plot the steady state microwave and mechanical populations as a function of the field strength by \cref{steady1,steady2,steady3}. From \cref{steady1}, we see that in the strong field limit ($g_r \gg 1$), $\theta_{ss} \rightarrow \pi/4$ or $-3\pi/4$, and from the other two equations, we see that the thermal populations balance to
\begin{gather}
n_c^{th,ss} = n_m^{th,ss} \rightarrow \bar{n}_b - \zeta\frac{\Delta n_b}{2}. \label{strongCool}
\end{gather}
In this limit, the total populations also converge to this value, so the instantaneous state is essentially a thermal state with a loss rate that is a mixture of the microwave and mechanical loss rates. From Eq. (\ref{strongCool}), we see that the mechanical population cannot be reduced below the equilibrium thermal population of the microwave bath, so if the frequency of the microwave cavity is low, ambient cooling is required if significant optomechanical cooling is to occur. In typical systems, where $\omega_c \gg \Omega_m$ and thus $n_c^b \ll n_m^b$, if the cavity loss rate is much faster than the mechanical one, the interaction will push the mechanical population towards zero, while equal loss rates will drive the population to $n_m^b/2$, and if the cavity loss rate is much slower than the mechanical one, it will fail reduce the mechanical population significantly below $n_m^b$.

Our results are in qualitative agreement with the experimental results of \cite{teufelCooling}, where it was found that the cooling increases with the field strength, but reaches a limit for a large field. From \cref{drivevss}, we see that there is no further reduction in the steady state temperature (thermal phonon number) for the mechanical mode if the pump field is increased above $g_r \approx 10$. We also see that the large-pump population limit is determined by the relative loss rate $\zeta$.

 %In the typical optomechanical case, $n_m^b \gg n_c^b$. %, so the large-pump limit population will reach $n_m^b(1 - \zeta)/2$ which agrees with the steady-state occupancy limit found in \cite{PhysRevLett.101.263602}. We note however that when applying this to electronic cavities with frequencies that are of the same order as the mechanical mode frequency, the cavity bath population must be considered. 

In the case where the microwave mode frequency is much larger than the mechanical mode frequency, then setting $n_c^b \rightarrow 0$, we obtain the following equations from \cref{steady1,steady2,steady3}:
\begin{gather}
\Delta n_{th}^{ss} \rightarrow \frac{\sqrt{1 + g_r^2} (1 - \zeta^2)n_m^b}{1 + g_r^2 - \zeta^2},\label{splitSSdiff}\\
\bar{n}_{th}^{ss} \rightarrow \frac{n_m^b}{2} \frac{(1 - \zeta)(1 + g_r^2 + \zeta)}{1 + g_r^2 - \zeta^2}.\label{splitSSavg}
\end{gather}

\begin{figure}[ht]
\includegraphics[width=\columnwidth]{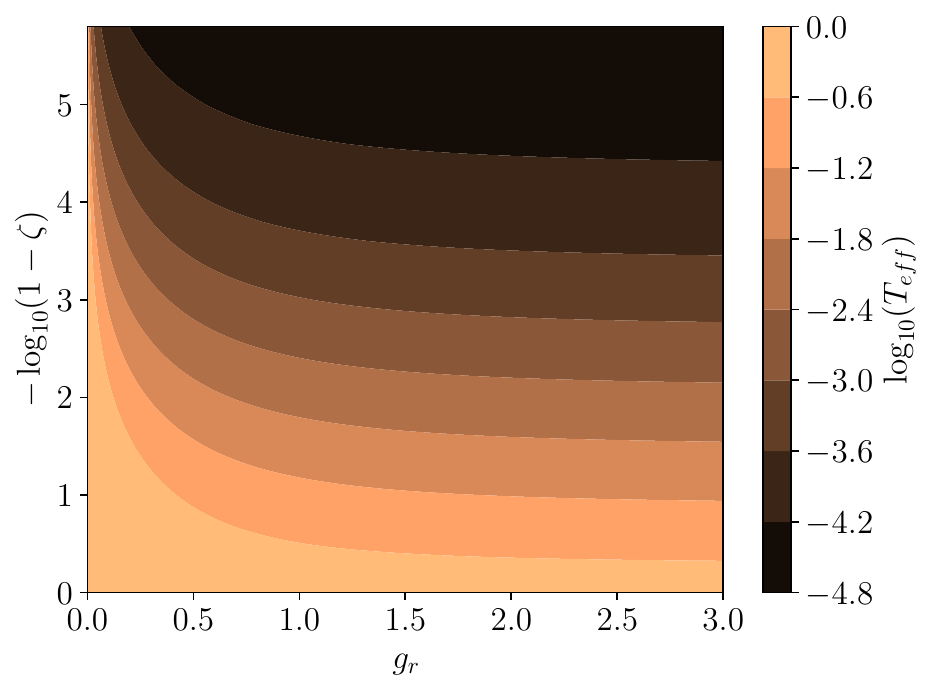}
\caption{Steady state mechanical mode temperature as a function of relative loss rate $\zeta$ and field strength $g_r$. The temperature scale is logarithmic, as is the scale for $\zeta$, where 0 corresponds to matched loss rates and 5 corresponds to $\zeta = 0.99999$. The system parameters used here are: $\Omega_m = 2\pi \times $10MHz, $T_b = 4\mathrm{K}$ ($n_m^b \approx 8\times 10^3$) and $\omega_c = 2\pi \times 100$GHz ($n_c^b\approx 0$).}
\label{heatmapfig}
\end{figure}

We can see how the relative loss and field strength cool the system by calculating this steady state value for a typical system. In \cref{heatmapfig}, we plot the logarithm of the effective temperature of the mechanical mode, $T_{eff} = \hbar\Omega_m/k\log(1 + 1/n_m^{th})$, determined by the dimensionless pump field and relative loss rates in a bath at liquid helium rather than cryostat temperatures. We see from this plot that a large field strength can only cool to a certain level, limited by the loss rates.

Cooling the system using the optomechanical process allows us to entangle the system using the red sideband interaction. However, given that the state is a thermal state with mixed populations, we should determine if the resulting state already exhibits entanglement between the microwave and mechanical modes. To do so, we consider the microwave and mechanical quadratures, $x_c$ and $x_m$, of the BTS. The rotated quadrature operator of the mechanical mode is defined as 
\begin{gather}
x_m(\beta_m) = \frac{b e^{i\beta_m} + b^\dag e^{-i\beta_m}}{2},
\end{gather}
where $\beta_m$ is a quadrature phase, and the microwave quadrature $x_c(\beta_c)$, is defined similarly. It is easily seen that for a BTS, this operator has $\expval{x_m} = 0$, and a variance given by $(\Delta x_m)^2 = (2n_m + 1)/4$. Thus, the state is not squeezed.
We can also consider the two-mode quadrature operators, 
\begin{gather}
X =  x_m(\beta_m) + x_c(\beta_c), \\
Y = x_m(\beta_m + \pi/2) - x_c(\beta_c + \pi/2)
\end{gather}
to determine the entanglement of the optomechanical system. Defining the correlation variance 
\begin{equation}\label{EqCorrVar}
    \Delta_{12}^2\equiv (\Delta X)^2 + (\Delta Y)^2,
\end{equation}
it is easy to show that for our beam-split thermal state, $\Delta_{12}^2 = n_m + n_c + 1 + \sin(2\theta)(n_c^{th} - n_m^{th})\sin(\phi_B + \beta_c - \beta_m)$, which is always greater than 1. The entanglement criterion for continuous variable systems is $\Delta_{12}^2 < 1$\cite{PhysRevLett.84.2726,PhysRevLett.84.2722}. Thus, modes that are initially in a thermal state will never be entangled by the beamsplitter interaction, as was shown in Kim \textit{et al.} \cite{PhysRevA.65.032323}. However, by reducing the thermal population, we have made it possible to produce entanglement with the application of the blue sideband interaction, as is described in the next section. 

\section{Optomechanical Entanglement}\label{squeezing}
In this section, we consider the case where the pump field frequency is on the blue sideband ($\Delta \approx \Omega_m$) and the interaction Hamiltonian takes the form of \cref{squeezeint}, creating a parametric oscillation \cite{PhysRevA.93.023846}. 
In a previous work, we considered a similar system with non-degenerate spontaneous parametric down conversion into two optical modes that are initially at zero temperature (equivalent to $n_c^b = n_m^b = 0$). This process has the same interaction Hamiltonian that we are using here (Eq. (\ref{EqHprime}) with $H_I^{red}=0$). In that work we determined that the exact solution to \cref{master} is a two-mode STS \cite{PhysRevA.103.022418}. 
Additionally, in a recent work \cite{PhysRevA.110.063718} we showed that when the interaction with a thermal environment is included in the single-mode case, the exact solution to the LME is a STS. In this section, we show that even for non-zero temperatures, the exact solution to the LME for the case of the blue sideband interaction Hamiltonian of \cref{squeezeint} is a two-mode STS that depends, in general, on the bath temperature and the difference between loss and the effective thermal populations of the two modes.

%For a two-mode STS to be the state of our system under the interaction, 
We again assume that the optomechanical system is initially in a two-mode thermal state or two-mode squeezed thermal state. To show that such a state will remain a two-mode squeezed thermal state for all time, we start by writing the density operator as 
\begin{gather}
\rho(t) = S(\xi(t))\rho_T(n_c^{th}(t), n_m^{th}(t))S^\dag(\xi(t)),
\label{STS}
\end{gather}
where
\begin{equation}
S(\xi) = \exp\left[ (\xi^* d b - \xi d^\dag b^\dag)\right] \label{squeezeOP}
\end{equation} 
is the two mode squeezing operator, with the time-dependent, complex squeezing factor $\xi(t) = u(t)e^{i\phi_S(t)}$ and the thermal state density operator is the same as that defined in the previous section.

In Appendix B, we show that the solution to the LME is indeed the two mode squeezed thermal state with thermal populations, squeezing amplitude, and phase that obeys the coupled ordinary differential equations
\begin{align}
\frac{dn_c^{th}}{dt} = &\kappa c^2 (n^b_c - n_c^{th}) + \Gamma_m s^2(n_m^b + n_m^{th} + 1), \\
\frac{dn_m^{th}}{dt} = &\Gamma_m c^2 (n^b_m - n_m^{th}) + \kappa s^2(n_c^b + n_m^{th} + 1), \\
2\frac{du}{dt} = &-i \gamma_0(\bar\alpha e^{-i\phi_S} - \bar\alpha^* e^{i\phi_S}) - \frac{cs}{n_c^{th} + n_m^{th} + 1} \nonumber \\ &\times \left( \begin{aligned} &\kappa(2n_c^b + 1) + \Gamma_m (2n_m^b + 1) \\ &+ (n_m^{th} - n_c^{th})(\kappa - \Gamma_m) \end{aligned} \right), \label{u0} \\
\frac{d\phi_S}{dt} = &- \gamma_0\frac{\bar\alpha e^{-i\phi_S} + \bar\alpha^* e^{i\phi_S}}{\tanh(2u)} + \Delta - \Omega_m,\label{phi0}
\end{align}
where $c \equiv \cosh(u)$ and $s \equiv \sinh(u)$.
In what follows, we again consider the pump field to be unchirped, such that it takes the form $\bar\alpha(t) = \alpha_0(t) e^{i\phi_L}$ for real $\alpha_0(t)$ and constant $\phi_L$. We begin the state in an unsqueezed state ($u(0) = 0$), and to prevent divergence in \cref{phi0}, we require $\bar\alpha(0) e^{-i\phi_S(0)} + \bar\alpha(0)^* e^{i\phi_S(0)} = 0$; thus we choose the initial phase to be $\phi_S(0) = \phi_L - \pi/2$. %We can see the phase must now evolve linearly with the pump frequency to be $\phi_S(t) = \phi_L - \pi/2 + \Delta_- t + \sigma(t)$, where $\Delta_- = \Delta - \Omega_m$ is the detuning from resonance. 
In what follows, we will only consider the case without detuning ($\Delta = \Omega_m$), where the phase is given simply by $\phi_S = \phi_L - \pi/2$ for all time. We note that adding detuning will reduce the magnitude of the squeezing operator $u$, similar to how detuning from the red sideband limits the mixing angle. 

We see that under the above conditions, the first term in \cref{u0} is simplified to $2\gamma_0|\alpha_0|$. We now define the dimensionless pump field strength $g_b = 2 \gamma_0 |\alpha_0|/\Gamma_+$. This can depend on time, but in what follows we will consider $g_b$ to be time-independent after it is turned on at time $t=0$. 
With these definitions and using the dimensionless time $\tilde{t}$, the dynamic equations become
\begin{align}
\frac{dn_c^{th}}{d\tilde{t}} &= (1 + \zeta)c^2(n_c^b - n_c^{th}) + (1 - \zeta)s^2(n_m^b + n_c^{th} + 1), \\
\frac{dn_m}{d\tilde{t}} &= (1 - \zeta)c^2(n_m^b - n_m^{th}) + (1 + \zeta)s^2(n_c^b + n_m + 1), \\
\frac{du}{d\tilde{t}} &= \frac{g_b}{2} - \frac{cs}{n_c^{th} + n_m^{th} + 1}\left( \begin{aligned} &n_c^b + n_m^b + 1 + \\ \zeta(&n_c^b - n_c^{th} + n_m^{th} - n_m^b) \end{aligned} \right).
\end{align}
We can further simplify these results using \cref{bars,deltas} to obtain 
\begin{align}
&\begin{aligned} 2\frac{d\bar{n}_{th}}{d\tilde{t}} = &\cosh(2u)\left[(2\bar{n_b} + 1)  + \zeta(\Delta n_{th} - \Delta n_b)\right] \\ &- (2\bar{n}_{th} + 1), \end{aligned} \label{nbar} \\
& \begin{aligned} \frac{d\Delta n_{th}}{d\tilde{t}} = &(\Delta n_b - \Delta n_{th}) - \zeta (2\bar{n}_b + 1) \\ &+ (2\bar{n}_{th} + 1) \zeta \cosh(2u), \end{aligned} \label{deltan} \\
&\frac{du}{d\tilde{t}} = \frac{g_b}{2} - \frac{cs}{2\bar{n}_{th} + 1} \left( 2\bar{n}_b + 1 + \zeta (\Delta n_{th} - \Delta n_b) \right). \label{u2}
\end{align}

In contrast to a single-mode squeezing operator, the two mode operator does not reduce the uncertainty in a quadrature of an individual mode below the vacuum level, but it can reduce the uncertainty of a combination of quadrature operators in the two modes. We quantify this using the correlation variance between the modes $\Delta_{12}^2$ that was defined in \cref{EqCorrVar}. For the two-mode squeezed thermal state, this correlation variance is given simply by $\Delta_{12}^2 = (n^{th}_1 + n^{th}_2 + 1)\exp(-2u)$ \cite{PhysRevA.103.022418}. Moreover, one can show that \cref{nbar,deltan,u2} can be used to derive the following  evolution equation for the correlation variance:
\begin{gather}
\frac{d \Delta_{12}^2}{d\tilde{t}} = (2\bar{n}_b + 1) + \zeta(\Delta n_{th} - \Delta n_b) - (1 + g_b) \Delta_{12}^2. \label{corr}
\end{gather}
%Alternatively, if the detectors measuring the fields of the two modes are out of phase by $\pi/2$, they will measure the anti-correlation variance, which takes the same form as the correlation variance but with $g_b \rightarrow -g_b$. 
In the limit that the two modes are identical, \cref{corr} reduces to the form of the squeezed single-mode quadrature found in Ref. \cite{PhysRevA.110.063718}.

We can also find the following steady state solutions for this system by setting the time derivatives in \cref{deltan,u2,nbar,corr} to zero:
\begin{gather}
u_{ss} = \frac{1}{2}\tanh^{-1}(g_b), \label{uss} \\
2\bar{n}_{th}^{ss} + 1 = (2\bar{n}_b + 1)\frac{(1 - \zeta^2)\sqrt{1 - g_b^2}}{1 - g_b^2 - \zeta^2} \label{nbarss}, \\
\Delta n_{th}^{ss} - \Delta n_b = (2\bar{n}_b + 1)\frac{g^2\zeta}{1 - g_b^2 - \zeta^2} \label{deltanss}, \\
\Delta_{12,ss}^2 = (2\bar{n}_b + 1) \frac{(1 - g_b)(1 - \zeta^2)}{1 - g_b^2 - \zeta^2}. \label{corrss}
\end{gather}
As Vendromin and Dignam  have shown \cite{PhysRevA.103.022418}, the system will only reach a steady-state value if the field strength is limited such that $g_b^2 < 1 - \zeta^2$. Despite this, using a pulsed $g_b(t)$ above this limit can reduce the correlation variance below the steady-state value for a short time, as we will discuss in the next section. %before it rises well above it due to the loss of entangled pairs. %, at which time the pulse can end.

We can also rewrite \cref{nbar,deltan,u2} without explicit reference to the bath population. Defining new variables %$\bar{n}_{th}^0$, $\Delta n_{th}^0$, and $\tilde{\Delta}_{12}^2$ such that
\begin{gather}
\bar{n}_{th}^0 = \frac 12 \left(\frac{2\bar{n}_{th} + 1}{2\bar{n}_b + 1} -1 \right), \label{scale1} \\
\Delta n_{th}^0 = \frac{\Delta n_{th} - \Delta n_b}{2\bar{n}_b + 1} \label{scale2} \\
\tilde{\Delta}_{12}^2 = \frac{\Delta_{12}^2}{2\bar{n}_b + 1}, \label{scale3}
\end{gather}
we obtain the dynamic equations:
\begin{gather}
2\frac{d \bar{n}_{th}^0}{d\tilde{t}} = \cosh(2u)(1 + \zeta \Delta n^0_{th}) - (2\bar{n}_{th}^0 + 1), \label{nbarbase} \\
\frac{d \Delta n_{th}^0}{d\tilde{t}} = (2\bar{n}_{th}^0 + 1)\cosh(2u) - \zeta - \Delta n_{th}^0, \label{deltanbase} \\
2\frac{du}{d\tilde{t}} = g_b - \frac{\sinh(2u)}{2\bar{n}_{th}^0 + 1} (1 + \zeta \Delta n_{th}^0), \label{ubase} \\
\frac{d\tilde{\Delta}_{12}^2}{d\tilde{t}} = 1 + \zeta \Delta n_{th}^0 - (g_b + 1)\tilde{\Delta}_{12}^2. \label{corrTilde}
\end{gather}
These equations are independent of the equilibrium bath temperature for the initial conditions $n_c^{th} = n_c^b, n_m^{th} = n_m^b$. Therefore, for a given $g_b(t)$, we can use these equations to calculate the state evolution from thermal equilibrium regardless of the temperature of the bath. %Additionally, it is convenient to use these scaled variables even when the state begins outside of thermal equilibrium, to keep computed variables small. 

% This allows for the calculation of the system dynamics for any bath population when the pump strength is known, as the state variables will only differ by the scaling factor for the equilibrium initial conditions. This transformation allows for easier calculation of a system when beginning it outside of the typical initial conditions when the bath population is very large, as is the case for optomechanical setups.

\section{Large transient entanglement scheme} \label{scheme}
In Sec. \ref{cooling}, we showed that the optomechanical interaction, with a pump cavity field on the red sideband, can cool the mechanical oscillator to an occupancy much below that of the environment. Additionally, in Sec. \ref{squeezing} we showed that the blue-sideband interaction can entangle the microwave and mechanical modes. %, dependent on the bath population. %thermal occupancy degrades the level of entanglement generated from the pure blue-sideband optomechanical interaction. 
In this section, we bring these two results together, presenting a scheme where the system first interacts with a pump field in the red-sideband, then is allowed to relax to a non-equilibrium thermal state where the mechanical system is cooled. Finally a blue sideband field is introduced to generate entangled photon-phonon pairs. Such a scheme is similar to the experiment of Palomaki \textit{et al.} \cite{doi:10.1126/science.1244563}, where they finally converted the entangled phonons back into photons using the red sideband interaction. %A similar scheme has been discussed using the Langevin equations with pulsed pumping \cite{PhysRevA.84.052327}, which was implemented in \cite{doi:10.1126/science.1244563}.

To begin, we need to determine what kind of state we will have after the beamsplitter interaction ends. To that end, we reexamine the beamsplitter state equations of motion when the field is removed by setting $g_r \rightarrow 0$. %Turning the pump off, we can see from \cref{thetadimless} how the mixing angle evolves. 
We will begin by assuming the populations will not decay as fast as the mixing angle $\theta$ does, and so take the values of the populations to be the steady state values for a constant red-sideband field determined previously in \cref{splitSSdiff,splitSSavg}. The time evolution of $\theta$ is then given by 
\begin{gather}
\frac{d\theta}{d\tilde{t}} = - \gamma\sin(2\theta), \label{angleDecay}
\end{gather}
where, from \cref{thetadimless}, the decay rate is given by 
\begin{align}
\gamma &= \sqrt{1 + {g_r'}^2}, \label{gammaApprox}
\end{align}
where $g_r'$ is the steady state field prior to setting $g_r \rightarrow 0$. We can solve \cref{angleDecay} to obtain the time evolution for $\theta$: 
\begin{gather}
\tan\theta = \frac{\gamma - 1}{\sqrt{\gamma^2 - 1}}e^{-2\gamma \tilde t},
\end{gather}
Using \cref{mechpopdimless}, we can also determine the rate of rethermalization for the mechanical mode; if we assume, as usual, that $n_m^b \gg n_c^b$, then initially, for large $g_r'$, 
\begin{gather}
\frac{dn_m^{th}}{d\tilde t} \approx \frac{1-\zeta}{2}n_m^b - n_m^{th},
\end{gather}
and as the $\theta$ decays to zero the rate becomes $dn_m^{th}/d\tilde t = (1 - \zeta)(n_m^b - n_m^{th})$. Both of these rates will be much slower than the decay rate for $\theta$ for $g_r'\gtrsim 5$. %Additionally, $\zeta \rightarrow 1$ ($\kappa \gg \Gamma_m$) also slows the rethermalization of the mechanical population to thermal equilibrium. 
Thus, our initial assumption that the mixing angle will decay very rapidly compared to the timescale on which the mechanical mode will rethermalize is well justified. 
%We can see that the rate $\gamma$ at which $\theta$ decays is significantly faster than the rate the mechanical or microwave mode will reach thermal equilibrium, $\propto (n_m^{th} - n_m^b) \Gamma_m/\kappa $ and $\propto (n_c^{th} - n_c^b)$ respectively, so long as $\zeta$ is close to 1. 

Now, with the mechanical mode cooled and in a thermal state, we activate the blue sideband pump to entangle the modes. We saw in Sec. IV that the steady state entanglement condition limits the field strength $g_b$. Under that limit, the correlation variance will only be reduced by $(1+\zeta)/2$, which is negligible for our purposes due to the large environmental thermal population. We therefore do not attempt to achieve steady state two-mode squeezing and instead we examine the short-term entanglement found using $g_b$ values that are above threshold value, $g_b^{th}\equiv \sqrt{1 - \zeta^2}$. %In the steady-state, the system will rethermalize and entanglement will be lost, so we should consider the early time entanglement enabled by pumping with $g_b^2 > 1 - \zeta^2$, rather than aiming for a steady-state solution. 
In the case where the initial state is in equilibrium with the environment, $g_b> g_b^{th}$ would cause the runaway generation of thermal photons, making entanglement impossible. 
However, for a system in which the less lossy mode is below its equilibrium temperature, we can use a strong field to create entangled pairs for a time duration less than the rethermalization time. %without losing many to the thermal environment. 
This can be seen by considering \cref{corr}. The cooled system will initially have $\Delta n_{th} \ll \Delta n_b$ and $\zeta > 0$. In the case of $n_c^b \approx 0$, the correlation variance will evolve as $d\Delta_{12}^2/d\tilde t = 1 + n_m^b (1-\zeta) - (1 + g_b)\Delta_{12}^2$. We note that this is exactly the equation for quadrature squeezing in degenerate SPDC, but with a reduced initial bath population of $n_m^b(1-\zeta)$, which thus enables significant entanglement, as we shall now show.
%We can see this in \cref{corrmin}, where as $\zeta \rightarrow 1$ and near zero $\Delta n_{th}$ will greatly lower the minimum correlation variance. 

We consider both the minimum possible correlation variance and the amount of time the states are entangled. %, \textit{i.e.}, $\Delta_{12}^2 < 1$, to evaluate the entanglement of the modes.
Going forward, we use  \cref{nbar,deltan,u2} to simulate the entanglement process, with the system in a thermal state, with the initial thermal populations given by \cref{strongCool}. In addition, we set the bath populations to be $n_m^b = 75, n_c^b = 0$, corresponding to a mechanical oscillator with resonant frequency of $\Omega_m \approx 2\pi \times 10$MHz, a microwave mode with frequency $\omega_c \approx 2\pi \times 100$GHz, and a cryogenic temperature of $T \approx$ 36mK. For significantly hotter temperature of say 4K, the results are qualitatively similar, but value of $\zeta$ closer to one is required to achieve entanglement.  %We consider various relative loss rates and consider the entanglement characteristics for varying squeezing blue sideband pump strengths. 
% To use the scaled equations in a computational simulation, the cooled populations of \cref{splitSSdiff,splitSSavg} are used in \cref{scale1,scale2} to determine the initial conditions, which produce
% \begin{gather} 
% \bar{n}_{th}^0(0) = \frac{-\zeta n_m^b}{2(n_m^b + 1)}, \\
% \Delta n_{th}^0(0) = -\frac{n_m^b}{1 + n_m^b}.
% \end{gather}
% Note that the system is entanglement if the scaled correlation variance satisfies $\tilde{\Delta}_{12}^2 < 1/(n_m^b + 1)$.

\begin{figure}[ht]
\includegraphics[width=\columnwidth]{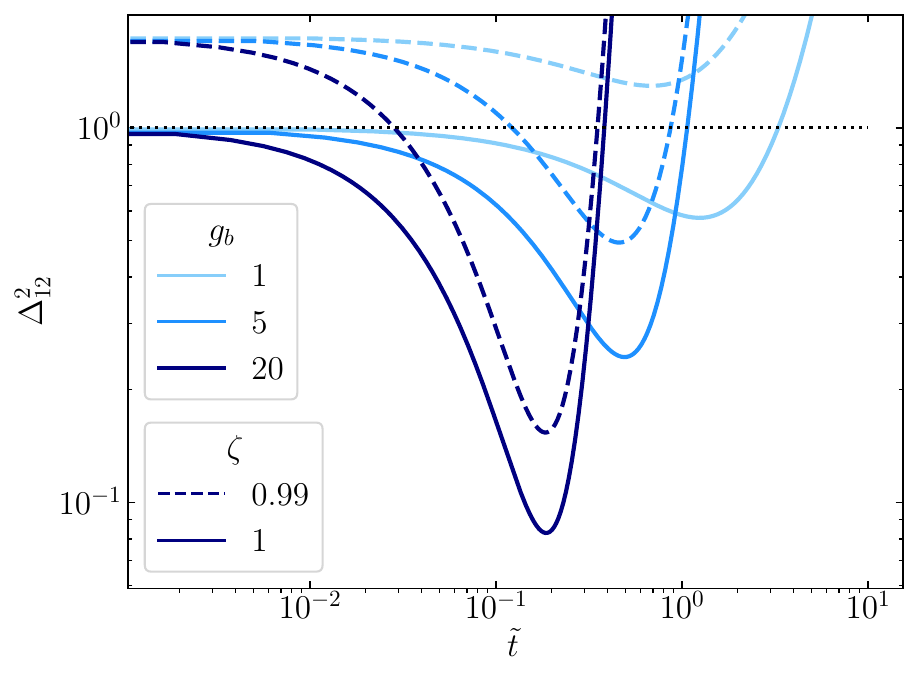}
\caption{Correlation variance ${\Delta_{12}^2}$ as a function of time in the large cooling limit for two different relative loss rates $\zeta$ and three different blue sideband field strengths. The dotted line at $\Delta_{12}^2 = 1$ indicates the minimum correlation variance entanglement threshold.}
\label{3corvar}
\end{figure}

\begin{figure}[ht]
\includegraphics[width=\columnwidth]{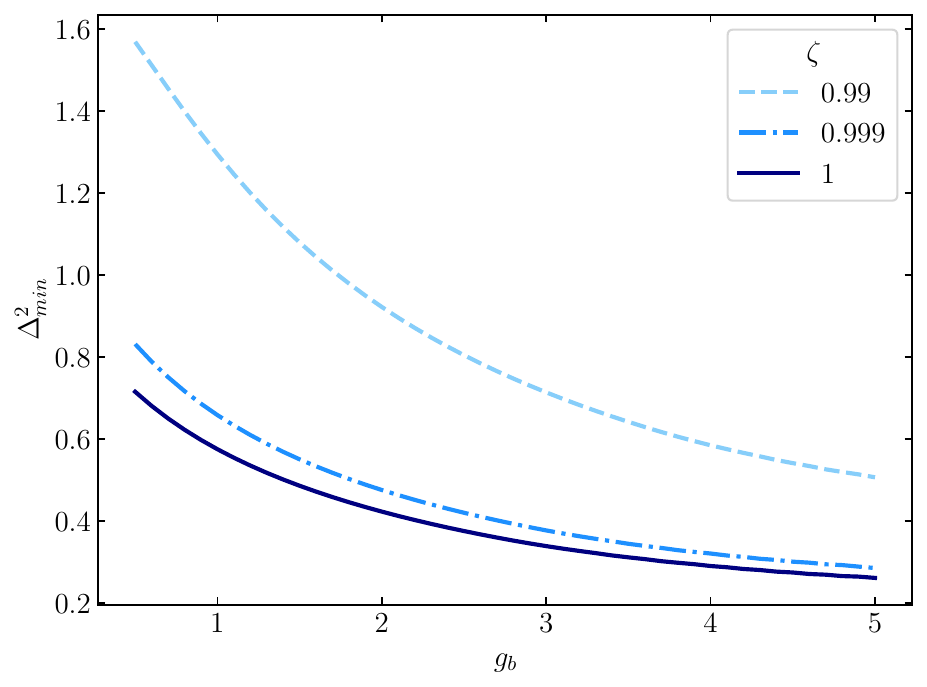}
\caption{Correlation variance ${\Delta_{12}^2}$ minimums as a function of $g_b$ in the large cooling limit for three different relative loss rates $\zeta$.}
\label{mins}
\end{figure}

In \cref{3corvar}, we plot the correlation variance for different values of $g_b$ and $\zeta$ as a function of time, where the initial state is cooled. We see that the blue sideband field creates transient entanglement. % prior to loss pushing the correlation variance above the entanglement threshold. 
We also see that for the optimal case that $\zeta \rightarrow 1$ (solid lines) weaker field can result in entanglement, but a larger field strength is required to achieve lower correlation variance. %We also note that the anti-correlation variance will also be increased faster by stronger pumps, but is decreased by the better relative loss rate $\zeta \rightarrow 1$. 
To see the impact of the loss ratio $\zeta$ and the field strength $g_b$ on the entanglement, in \cref{mins} we plot the minimum of the correlation variance ($\Delta_{min}^2$) as a function of the field strength. As can be seen, the smallest correlation variance is achieved for large $g_b$ and $\zeta$ approaching 1. %It is possible to investigate the same behaviour using a time dependent pulse for $g_b(t)$, though this will not result in a longer lasting entanglement for the same peak field strength.

% We next consider the minimum correlation variance to occur at a time $t_1$. For values of $\zeta \rightarrow 1$, the minimum correlation variance occurs early in the pumping when $\Delta n_{th}^0(t_1) \approx \Delta n_{th}(0) \approx -1$, so 
% \begin{gather}
% \tilde{\Delta}_{12, min}^2 \approx \frac{1 - \zeta}{1 + g_{b}}. \label{mindelta}
% \end{gather}
% In \cref{mins}, we show how the minimum correlation variance is reduced by a stronger blue pump.

\begin{figure}[ht]
\includegraphics[width=\columnwidth]{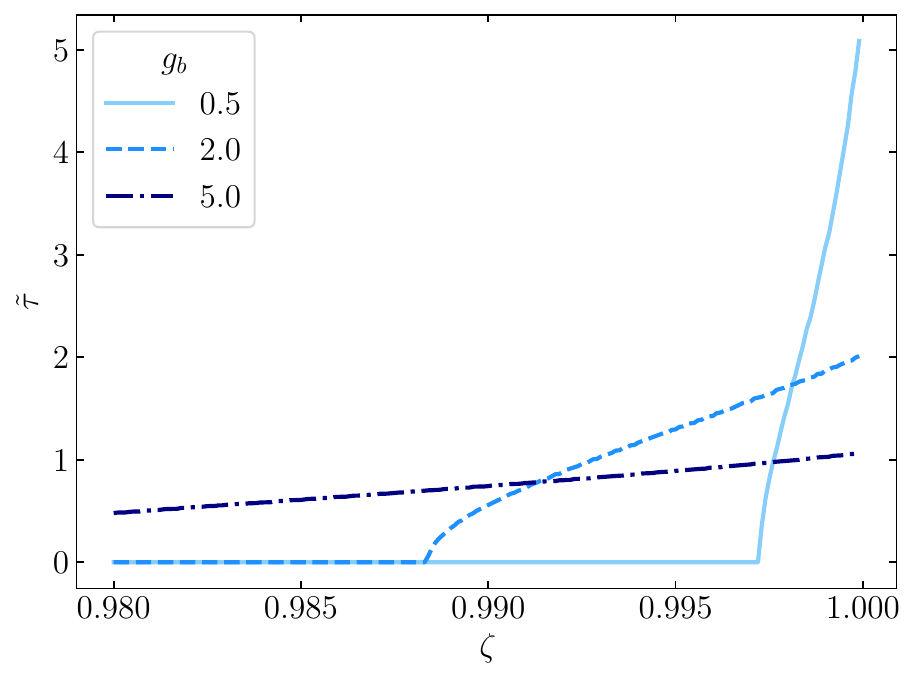}
\caption{Time, $\tilde\tau$ below the entanglement threshold ($\Delta_{12}^2 =1$) as a function of $\zeta$ for different values of $g_b$ for a system initially in the large cooling red sideband limit.% We use three mean field strength values, $g_b$, which shows that weaker fields require larger $\zeta$ to achieve optomechanical entanglement, though the entanglement time can be significantly longer for weak fields at larger $\zeta$.
}
\label{entTime}
\end{figure}

Due to runaway thermalization, although large field strengths lower the correlation variance more, they also break the entanglement threshold faster than smaller fields. Thus, we now consider the time duration over which the correlation variance is below threshold. In \cref{entTime}, we plot $\tilde\tau$, the time the system is entangled, against $\zeta$, for different pump strengths. We see that for larger values of $\zeta$, small pump strengths are able to create entanglement for longer than strong pump strengths, although again, large pumps are required to entangle systems with low $\zeta$. 

% From the above discussion we see that we can thus maximize the time over which the system is entangled for a given set of system parameters %in a thermal environment 
% by maximizing the relative loss ratio $\zeta$, cooling the mechanical system with a strong red-sideband field, and then activating the blue-sideband field with a strength that is slightly greater than the minimum required for entanglement. 
% That is, we choose
% \begin{gather}
% g_{b} > n_m^b (1 - \zeta) - \zeta.
% \end{gather}
% With non-optimized loss rates, stronger fields will often create longer lasting entanglement. 
The correlation variance increases due to the thermal populations arising from the blue sideband pump in conjunction with loss. One may be tempted to mitigate this effect by removing the pump ($g_b = 0$) after the system is entangled to a desirable level. % and allowing the system to enter free evolution. 
However, cutting off the pump when the system is below the entanglement threshold only breaks entanglement sooner, as the loss to the environment is not mitigated by further entangled-pair production. %Thus, system free evolution, unlike allowing a constant strength mean field to remain after optomechanical entanglement, will allow the system to settle into thermal equilibrium at a significantly larger correlation variance.

\begin{figure}[ht]
\includegraphics[width=\columnwidth]{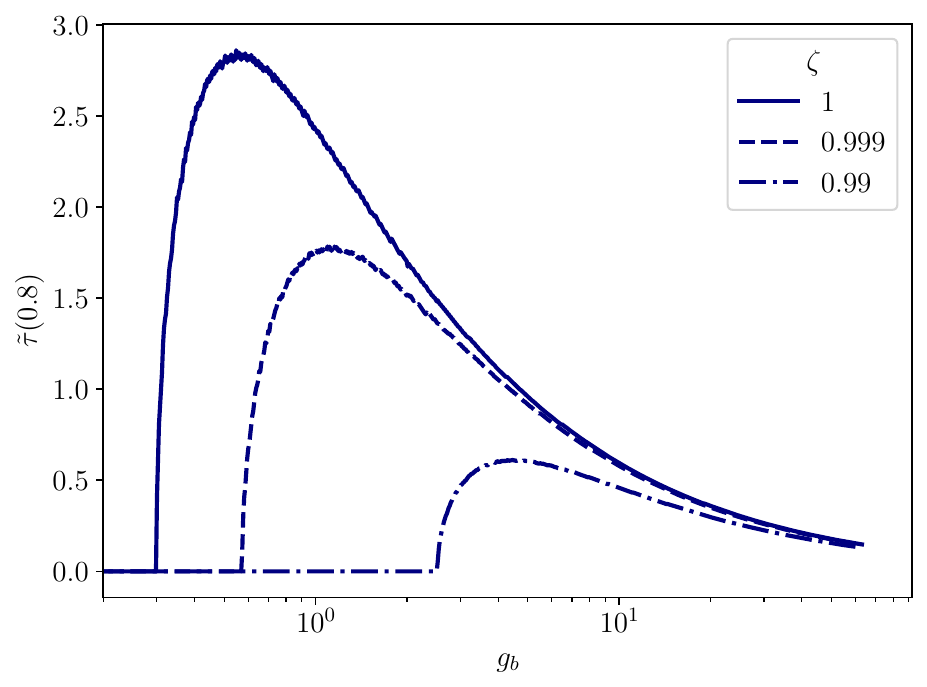}
\caption{Time over which the correlation variance satisfies $\Delta_{12}^2 < 0.8$, as a function of the blue sideband pump strength $g_b$ for three different relative loss rates, $\zeta$.}
\label{timeunder}
\end{figure}

Rather than simply trying to keep the system entangled for as long as possible, we may instead desire that the correlation variance stay below a specified value for as long as possible, so as to keep our entanglement robust to other sources of loss arising in transmission, or from other components in a composite system. %desire a specified level of correlation variance to last as long as possible. 
%If this is the case, we want to keep the correlation variance below some specified value $\Delta_t$ for a long as possible. 
%We wish to find the optimal field strength to keep the correlation variance below that threshold for as long as possible. 
For instance, consider that we wish to reduce the correlation variance below 0.8, and need to choose a pump strength to maximize the time under this variance. In \cref{timeunder}, we plot this time $\tilde\tau(0.8)$ as a function of pump strength for different values of $\zeta$. We see that, depending on the specific value of $\zeta$, different pump strengths lead to the longest-lasting correlation variance under the target value of 0.8.

\begin{figure}[ht]
\includegraphics[width=\columnwidth]{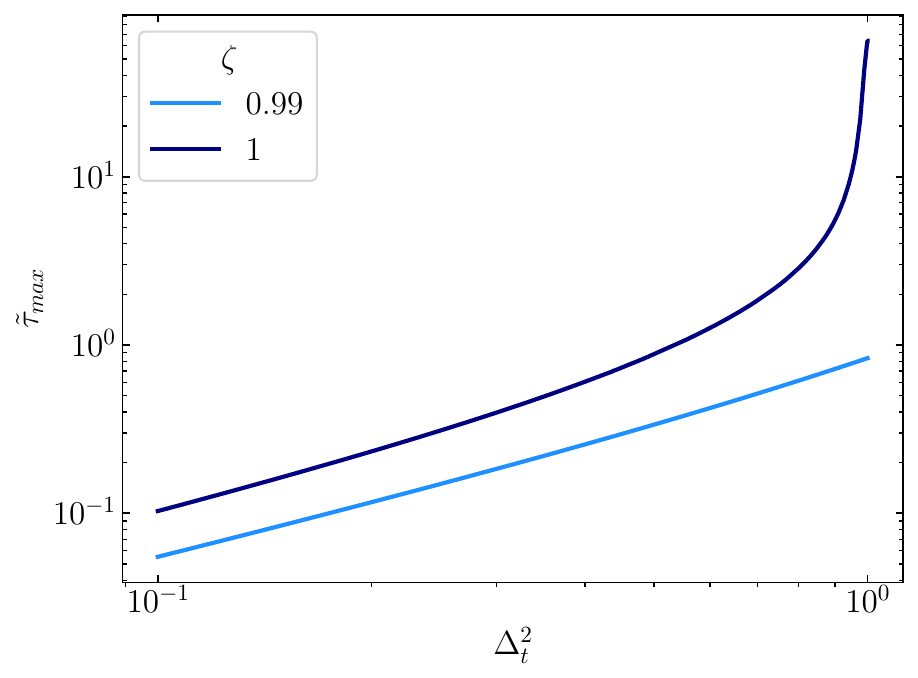}
\caption{Maximum time below threshold as a function of threshold value for two relative loss rates $\zeta$, when the optimal field strength is used (see the peak values in \cref{entTime} for $\Delta^2_t=0.8$).}
\label{optimalTime}
\end{figure}

We can also see that \cref{timeunder} has a maximum value of time under the target correlation variance $\tilde \tau_{max}$ at an optimal pump strength $g_{opt}$.%, for which the time under the target correlation variance is maximized ($\tilde\tau_{max}$) is the optimal pump strength for the target level. 
This maximum time and optimal pumping strength depends on the correlation variance limit desired and the loss parameter, $\zeta$. A lower limit requires a larger pump and results in a shorter duration. In \cref{optimalTime}, we plot the maximum time achieved ($\tilde \tau_{max}$) as a function of the correlation variance target. %\cref{optimalPump,optimalTime}, we plot the optimal pump strength ($g_{opt}$), and the maximum time achieved ($\tilde\tau_{max}$), as a function of the correlation variance target. As can be seen in \cref{optimalTime}, 
This time is significantly reduced when the target value is reduced, but improves for $\zeta \rightarrow 1$. 

\begin{figure}[ht]
\includegraphics[width=\columnwidth]{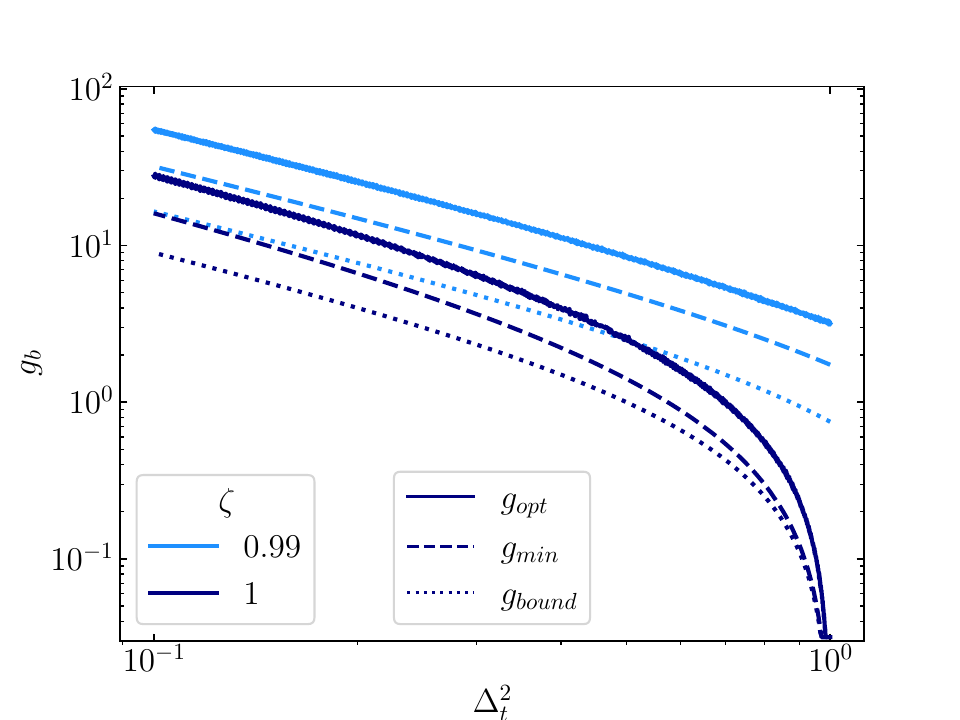}
\caption{Optimal field strength as a function of the target correlation variance $\Delta^2_t$ for different relative loss rates $\zeta$. The solid lines indicate $g_{opt}$, the field that maximizes the time under the correlation variance target $\Delta^2_t$ for two different $\zeta$ values. The dashed lines indicate $g_{min}$ the minimum field strength required to reduce $\Delta_{12}^2$ below $\Delta^2_t$ for any amount of time. The dotted lines indicate the analytic lower bound to $g_{min}$, given by $g_{bound} = [(1-\zeta)n_b^m + 1]/\Delta^2_t + 1$.}
\label{optimalPump}
\end{figure}

The field strengths ($g_{opt}$) that produce these maximum times are plotted in \cref{optimalPump}, where we also show the minimum field strength ($g_{min}$) required to reach that correlation variance. %For each $\zeta$, the curves follow a similar pattern, with the optimal level being larger than the minimum by a scale factor for most correlation variance targets. 
To gain some additional insight into the dependence of the optimal field strength on system parameters, we set $\frac{d \Delta_{12}^2}{d\tilde{t}} =0$ in \cref{corr} to obtain the minimum value for $\Delta_{12}^2$. %The minimum field strength required to reach a target correlation variance will satisfy that relation, which assuming $n_c^b \approx 0$ is given by
Assuming $n_c^b \approx 0$, we obtain the following expression for the minimum field strength required to achieve a chosen target correlation variance:
\begin{gather}
g_{min} = \frac{(1-\zeta)n_b^m + 1 + \zeta\Delta n_{th}(t_1)}{\Delta^2_t} + 1,
\end{gather}
where $t_1$ is the time at which the correlation is a minimum and $\Delta^2_t$ is the target correlation variance. % reached at $t_1$.
This expression is not easily solved, as $\Delta n_{th}(t_1)$ depends on the target correlation. However, we can find a lower bound for this by approximating $\Delta n_{th}(t_1) \approx \Delta n_{th}(0) = 0$, which is valid in the large cooling limit. The resulting curve, $g_{bound} = [(1-\zeta)n_b^m + 1]/\Delta^2_t + 1$, is plotted in \cref{optimalPump} for two different values of $\zeta$. We see that the trends for these curves are similar to those for the computed optimal pump strength.
%, leading us to conclude that the optimal field strength follows a similar shape. 
For example, for $\zeta = 1$, the curve $g_{approx}(\Delta^2_t) \equiv 1.5 g_{bound}(\Delta^2_t)$ only differs from $g_{opt}(\Delta^2_t)$ by about 10\% over a range of over two orders of magnitude in $g_b$. The fit is similar for other values of $\zeta$, with different scaling factors, indicating that the general dependence of $g_{opt}$ on $\Delta^2_t$ is close to what is found for $g_{bound}$.

%In this work, we used a dimensionless timescale $\tilde t$. To get a sense of if our entanglement lasts for a usable duration, we should consider an example of a reasonable cavity ($\omega_c \approx 2\pi\times 10^{11}s^{-1}, \kappa \approx 10^5s^{-1}$). With this parameter, and $\kappa \gg \Gamma_m$, $\tilde t = 1$ unit corresponds to 10$\mu$s.

\section{Conclusion}
In this work, we found semi-analytic solutions to the Lindblad master equation for a lossy optomechanical system pumped in the red and blue mechanical sidebands of the cavity resonance. Using our solutions, we presented full state descriptions for each of these interactions. We then calculated the evolution during the laser-cooling and derived the pumping requirements to reach an entangled state based on the thermal occupation of the mechanical mode and the relative loss rates of the microwave and mechanical modes. We also derived a limit to the possible entanglement in an optomechanical system under a cooling-squeezing scheme based on the relative loss rate, squeezing pump strength, and environmental temperature. We furthermore showed that large pump strengths are needed to push the correlation variance below the entanglement threshold in an environment with a large thermal population, but that over pumping causes premature loss of all entanglement in these systems. This allowed us to determine the optimal pumping strength for optomechanical entanglement as a function of the desired entanglement threshold.

Our analysis has mainly focused on the results for negligible microwave bath populations. However, in the case where we employ low-frequency electronic cavities rather than microwave cavities, the cavity thermal populations may not be negligible, which can considerably reduce the entanglement potential of the optomechanical system. 

In this work, we used a dimensionless timescale $\tilde t=\Gamma_{+}t$ and found that entanglement could be maintained for $\tilde{\tau}$ in the range of 1 to 5. As a concrete example of a realistic system, consider a cavity with resonant frequency $\omega_c \approx 2\pi\times 10^{11}s^{-1}$ and $\kappa \approx 10^5s^{-1}$. For this cavity, assuming $\kappa \gg \Gamma_m$, $\tilde{\tau} = 1$ corresponds to a time of 10$\mu$s, which is a long time relative to typical measurement timescales.

Our results will help researchers working on optomechanical systems better understand the environmental precooling requirements to achieve specific levels of optomechanical entanglement, which could then be applied to other quantum resource generation. In future work, we hope to investigate the simultaneous red and blue sideband pumping of the optomechanical system in our formalism. This technique has been shown to produce squeezed mechanical states, which we interpret as the realization of simultaneous squeezed and beam-split thermal states.

\section*{Acknowledgements}
We thank Rob Knobel and Nathan Eddy for useful discussions. 
This work was supported by the Canada Foundation for Innovation and the Natural Sciences and Engineering Research Council of Canada (NSERC).

\setcounter{equation}{0}
\renewcommand{\theequation}{A\arabic{equation}}
\section*{Appendix A: Mechanical Cooling State}
In this section, we apply the method employed by Seifoory \textit{et al.} \cite{hossein} to show that the solution to the LME of \cref{master} with the Hamiltonian of \cref{EqHprime} is a beam-split two mode thermal state when $H_I^{blue} = 0$. In the derivation, we have two arbitrary modes with thermal populations denoted as $n_1$ and $n_2$, which correspond to the cavity and mechanical phonon thermal populations in the main text. We use the general form of the Hamiltonian
\begin{gather}
H = \hbar \omega_1 b_1^\dag b_1 + \hbar \omega_2 b_2^\dag b_2 + \hbar(\gamma b_1^\dag b_2 + \gamma^* b_1 b_2^\dag). \label{A1hamm}
\end{gather}
For later convenience, we will now define the operators
\begin{gather}
\sigma \equiv i(b_1^\dag b_2e^{i\phi_B} - b_1 b_2^\dag e^{-i\phi_B}), \\
\Omega \equiv (b_1^\dag b_2e^{i\phi_B} + b_1 b_2^\dag e^{-i\phi_B}).
\end{gather}

To see that the density operator that solves the LME for this Hamiltonian is a BTS, consider modifying \cref{STS} to take the form 
\begin{gather}
\rho(t) = B(\theta, \phi_B) \rho_T^{1/2}(n_1, n_2) O(t) \rho_T^{1/2}(n_1, n_2) B^\dag(\theta, \phi_B).
\end{gather}
Inverting this we obtain
\begin{gather}
O(t) = \rho_T^{-1/2}B^\dag\rho B\rho_T^{-1/2}.
\end{gather}
For the solution to hold, the operator $O(t)$ must be the identity for all times. Using the product rule and the LME, the operator's time derivative can be split into five parts:
\begin{gather}
\dot{O}(t) = \dot{O}_T(t) + \dot{O}_B(t) + \dot{O}_0 + \dot{O}_V + \dot{O}_L.
\label{odotB}
\end{gather}
The first term is defined using the thermal state change
\begin{gather}
\dot{O}_T = \frac{d\rho_T^{-1/2}}{dt} B^{\dag}\rho B \rho_T^{-1/2} + \rho_T^{-1/2} B^\dag \rho B \frac{d \rho_T^{-1/2}}{dt}
\end{gather}
or
\begin{gather}
\dot{O}_T = \acomm{J}{O}, 
\end{gather}
where
\begin{gather}
J \equiv \frac{d\rho_T^{-1/2}}{dt}\rho_T^{1/2} = \rho_T^{1/2}\frac{d\rho_T^{-1/2}}{dt}.
\end{gather}
For convenience, let $x_i \equiv e^{-\beta\hbar\omega_i}$. Then we obtain 
\begin{gather}
J = \frac{1}{2}\sum_{i=1}^2 \frac{1}{x_i}\frac{dx_i}{dt}(n_i - b_i^\dag b_i).
\end{gather}

The second term contains the time derivatives of the beamsplitter operator
\begin{align}
\begin{aligned}
\dot{O}_B(t) = \rho_T^{-1/2} \frac{dB^\dag}{dt} B \rho_T^{1/2} (\rho_T^{-1/2} B^\dag \rho B \rho_T^{-1/2}) \\+ (\rho_T^{-1/2} B^\dag \rho B \rho_T^{-1/2})\rho_T^{1/2} B^\dag \frac{dB}{dt}\rho_T^{-1/2},
\end{aligned}
\end{align}
which can be rewritten as
\begin{gather}
\dot{O}_B(t) = LO + OL^\dag, 
\end{gather}
where
\begin{align}
\begin{aligned}
L &\equiv -i\dot\theta (x_1^{-1/2}x_2^{1/2} b_1^\dag b_2 + x_1^{1/2}x_2^{-1/2}b_1 b_2^\dag) \\
& - \frac{\dot{\phi_B}}{2}(x_1^{-1/2}x_2^{1/2}b_1^\dag b_2 e^{i\phi_B} - x_1^{1/2}x_2^{-1/2}b_1 b_2^\dag e^{-i\phi_B})\sin(2\theta) \\
& - \frac{i}{2}\dot{\phi_B}(b_1^\dag b_1 - b_2^\dag b_2)(\cos(2\theta) - 1),
\end{aligned}
\end{align}
such that we can write $L = M + iN$.
The unperturbed Hamiltonian contributes
\begin{gather}
\dot{O}_0 = \rho_T^{-1/2} B^\dag \left( -i\comm{\omega_1 b_1^\dag b_1 + \omega_2 b_2^\dag b_2}{\rho} \right) B \rho_T^{-1/2},
\end{gather}
Which can be rewritten as
\begin{gather}
\dot{O}_0 = -i \left(\omega_1 G_1 O - \omega_1 OG_1^\dag + \omega_2 G_2 O - \omega_2 OG_2^\dag \right), 
\end{gather}
with
\begin{gather}
G_i \equiv \rho_T^{-1/2} B^\dag b_i^\dag b_i B \rho_T^{1/2}.
\end{gather} 
Again, these are split into commuting and anti-commuting parts $G_i = P_i + iQ_i$ so
\begin{gather}
\dot{O}_0 = -i\comm{\omega_1 P_1 + \omega_2 P_2}{O} + \acomm{\omega_1 Q_1 + \omega_2 Q_2}{O}.
\end{gather}
For $i, j = 1, 2$, we see 
\begin{align}
&\begin{aligned}
P_i = b_i^\dag b_i \cos^2\theta + &b_j^\dag b_j \sin^2\theta + \frac{i}{2}(x_i^{-1/2}x_j^{1/2} + x_i^{1/2} x_j^{-1/2}) \\ &\times (b_i^\dag b_j e^{\pm i\phi_B} - b_i b_j^\dag e^{\mp i\phi_B}) \cos\theta \sin\theta,
\end{aligned} \\
& \begin{aligned} Q_i = \frac{1}{2} &(x_i^{-1/2}x_j^{1/2} - x_i^{1/2}x_j^{-1/2}) \\ &\times(b_i^\dag b_j e^{\pm i\phi_B} + b_i b_j^\dag e^{\mp i\phi_B}) \cos\theta \sin\theta, \end{aligned}
\end{align}
where $i=1$ is ``+'' and $j=1$ is ``-''.

The perturbation, written as $V(t) = \hbar(\gamma b_1^\dag b_2 + \gamma^* b_1 b_2^\dag)$ contributes
\begin{gather}
\dot{O}_V = \rho_T^{-1/2}B^\dag \left( -\frac{i}{\hbar}\comm{V}{\rho} \right) B \rho_T^{-1/2}
\end{gather}
which can be written as a commuting and non-commuting part as
\begin{align}
&\dot{O}_V = -i\comm{\bar{P}}{O} + \acomm{\bar{Q}}{O}, \\
&\begin{aligned}
-2\bar{Q} \equiv (x_1^{-1/2}&x_2^{1/2} - x_1^{1/2}x_2^{-1/2}) \\ 
 \times \Big[&i\Omega(\cos^2\theta - \frac{1}{2})(\gamma e^{-i\phi_B} - \gamma^* e^{i\phi_B}) \\
& + \frac{\sigma}{2}(\gamma e^{-i\phi_B} + \gamma^* e^{i\phi_B}) \big],
\end{aligned}
\end{align} 
and we don't give the expression for $\bar{P}$ because it vanishes in the final calculation of $\dot{O}_V$.
%while $\bar{P}$ is irrelevant to the final calculation. 
Finally, the Lindblad terms contribute
\begin{widetext}
\begin{align}
\dot{O}_{L} = \sum_{i=1}^2 \left[
\begin{aligned}
\Gamma(n_i^b + 1) \Big( &\rho_T^{-1/2}B^\dag b_i \rho b_i^\dag B \rho_T^{-1/2} - \frac{1}{2}\rho_T^{-1/2} B^\dag (\acomm{b_i^\dag b_i}{\rho}) B \rho_T^{-1/2}\Big) \\
+ \Gamma n_i^b \Big( &\rho_T^{-1/2}B^\dag b_i^\dag \rho b_i B \rho_T^{-1/2} - \frac{1}{2}\rho_T^{-1/2}B^\dag (\acomm{b_i b_i^\dag}{\rho}) B \rho_T^{-1/2}\Big)
\end{aligned} \right].
\end{align}
\end{widetext}
Now, we define the operators
\begin{align}
& T_i \equiv \rho_T^{-1/2} B^\dag b_i B \rho_T^{1/2}, \\
& \tilde{T}_i \equiv \rho_T^{-1/2} B^\dag b_i^\dag B \rho_T^{1/2}, 
\end{align}
\begin{align}
& \begin{aligned}
\tilde{G}_i &\equiv \rho_T^{-1/2} B^\dag b_i b_i^\dag B \rho^{1/2} \\ &= G_i + \rho_T^{-1/2} B^\dag \comm{b_i}{b_i^\dag} B \rho_T^{1/2} \\ &= G_i + 1,
\end{aligned}
\end{align}
such that we can write
\begin{gather}
\rho_T^{-1/2}B^\dag b_i \rho b_i^\dag B \rho_T^{-1/2} = T_i O T_i^\dag, \\
\rho_T^{-1/2}B^\dag b_i^\dag \rho b_i^\dag B \rho_T^{-1/2} = \tilde{T}_i O {\tilde{T}_i}^\dag,
\end{gather}
to obtain
\begin{align}
\dot{O}_{L} &= \sum_{i=1}^2 \left[ \begin{aligned}
\Gamma(n_i^b + 1) &\left[T_i O T_i^\dag - \frac{1}{2}(G_i O + OG_i^\dag)\right] \\
+ \Gamma n_i^b &\left[\tilde{T}_i O \tilde{T}_i^\dag -\frac{1}{2}(\tilde{G}_i O + O\tilde{G}_i^\dag)\right] 
\end{aligned} \right] \nonumber \\
& = \sum_{i=1}^2 \left[
\begin{aligned}
\Gamma(n_i^b + 1) \left[T_i O T_i^\dag - \frac{1}{2}\acomm{P_i}{O} - \frac{i}{2}\comm{Q_i}{O}\right]& \\+ \Gamma n_i^b \left[\tilde{T}_i O \tilde{T}_i^\dag -\frac{1}{2}\acomm{P_i}{O} - \frac{i}{2}\comm{Q_i}{O} - O\right]& 
\end{aligned} \right].
\end{align}
Now, inserting the parts of the equation back into \cref{odotB}, and setting $O$ to the identity for all time yields
\begin{align}
\begin{aligned}
0 &= 2J + 2M + 2\omega_1 Q_1 + 2\omega_2 Q_2 + \frac{2}{\hbar}\bar{Q}  \\ &+ \sum_{i=1}^2 \left[\Gamma_i (n_i^b + 1)(T_iT_i^\dag - P_i) + \Gamma_i n_i^b (\tilde{T}_i \tilde{T}_i^\dag - P_i - 1) \right]. \label{simpleOB}
\end{aligned}
\end{align}
We note that we can write
\begin{align}
&\begin{aligned}
T_i T_i^\dag = &x_i b_i b_i^\dag \cos^2\theta + x_j b_j b_j^\dag \sin^2\theta \\ &+ ix_i^{1/2} x_j^{1/2}(b_i^\dag b_j e^{\pm i\phi_B} - b_i b_j^\dag e^{\mp i\phi_B})\sin\theta \cos\theta, 
\end{aligned} \\
&\begin{aligned}
\tilde{T}_i \tilde{T}_i^\dag = &x_i^{-1} b_i^\dag b_i \cos^2\theta + x_j^{-1} b_j^\dag b_j \sin^2\theta \\ &+ ix_i^{-1/2} x_j^{-1/2}(b_i^\dag b_j e^{\pm i\phi_B} - b_i b_j^\dag e^{\mp i\phi_B})\sin\theta \cos\theta.
\end{aligned}
\end{align}
which allow us to split the solved components of \cref{simpleOB} into linearly independent parts based on the mode operators.

First, from the terms proportional to $b_1^\dag b_1$, we have
\begin{align}
&\begin{aligned}
\frac{1}{x_1}&\frac{dx_1}{dt} = \Gamma_1(n_1^b + 1)(x_1 - 1)\cos^2\theta + \Gamma_1 n_1^b (x_1^{-1} - 1)\cos^2\theta \\
+ &\Gamma_2(n_2^b + 1) (x_1 - 1)\sin^2\theta + \Gamma_2 n_2^b (x_1^{-1} - 1)\sin^2\theta,
\end{aligned} \\
&\Rightarrow \begin{aligned}
\frac{dn_1}{dt} = \Gamma_1 [n_1^b - n_1^{th}]\cos^2\theta + \Gamma_2[n_2^b - n_1^{th}]\sin^2\theta.
\end{aligned}
\end{align}
Similarly, from the terms proportional to $b_2^\dag b_2$, we obtain
\begin{align}
\begin{aligned}
\frac{dn_2}{dt} = \Gamma_2 [n_2^b - n_2^{th}]\cos^2\theta + \Gamma_2[n_1^b - n_2^{th}]\sin^2\theta.
\end{aligned}
\end{align}
The terms proportional to $\sigma$ produce the following dynamic equation for $\theta$: 
\begin{align} \begin{aligned}
0 = & -\dot{\theta}(x_1^{-1/2}x_2^{1/2} - x_1^{1/2}x_2^{-1/2}) \\
+ & \frac{1}{2}(x_1^{-1/2}x_2^{1/2} - x_1^{1/2}x_2^{-1/2})(\gamma e^{-i\phi_B} + \gamma^* e^{i\phi_B}) \\
& \begin{aligned} + &\Gamma_1(n_1^b + 1) \cos\theta\sin\theta \\ &\times \big[ x_1^{1/2}x_2^{1/2} - \frac{1}{2}(x_1^{-1/2} x_2^{1/2} + x_1^{1/2}x_2^{-1/2})\big]\end{aligned} \\
&\begin{aligned} + &\Gamma_1 n_1^b \cos\theta\sin\theta \\ &\times \big[x_1^{-1/2} x_2^{-1/2} - \frac{1}{2}(x_1^{-1/2} x_2^{1/2} + x_1^{1/2}x_2^{-1/2})\big]\end{aligned} \\
& \begin{aligned} - &\Gamma_2(n_2^b + 1) \cos\theta\sin\theta \\ &\times \big[x_1^{1/2}x_2^{1/2} - \frac{1}{2}(x_1^{-1/2} x_2^{1/2} + x_1^{1/2}x_2^{-1/2})\big]\end{aligned} \\
&\begin{aligned} - &\Gamma_2 n_2^b \cos\theta\sin\theta \\ &\times \big[x_1^{-1/2} x_2^{-1/2} - \frac{1}{2}(x_1^{-1/2} x_2^{1/2} + x_1^{1/2}x_2^{-1/2})\big], \end{aligned}
\end{aligned} \end{align}
which can be written as 
\begin{align} \begin{aligned}
\frac{d\theta}{dt} = &+\frac{1}{2}(\gamma e^{-i\phi_B} + \gamma^* e^{i\phi_B})\\
& + \frac{\sin(2\theta)}{2}\Gamma_1\left(\frac{n_1^b}{n_2 - n_1} - \frac{n_2 + n_1}{2(n_2 - n_1)}\right) \\ &- \frac{\sin(2\theta)}{2}\Gamma_2\left(\frac{n_2^b}{n_2 - n_1} - \frac{n_2 + n_1}{2(n_2 - n_1)}\right).
\end{aligned} \end{align}
Finally, the terms proportional to $\Omega$ yields the equation
\begin{align}
\begin{aligned}
0 = (\omega_1 - \omega_2)&(x_1^{-1/2} x_2^{1/2} - x_1^{1/2}x_2^{-1/2})\cos\theta\sin\theta \\
- \frac{\dot{\phi_B}}{2}&(x_1^{-1/2} x_2^{1/2} - x_1^{1/2}x_2^{-1/2})\sin(2\theta) \\
- i(\gamma e^{-i\phi_B} - \gamma^* e^{i\phi_B})&(x_1^{-1/2} x_2^{1/2} - x_1^{1/2}x_2^{-1/2})(\cos^2\theta - 1/2),
\end{aligned}
\end{align}
which can be rewritten as
\begin{gather}
\frac{d\phi_B}{dt} = (\omega_1 - \omega_2) - i\frac{\gamma e^{-i\phi_B} - \gamma^* e^{i\phi_B}}{2 \tan(2\theta)}.
\end{gather}

Thus, we have shown that the solution to the LME for the Hamiltonian of \cref{A1hamm} is a BTS.

\setcounter{equation}{0}
\renewcommand{\theequation}{B\arabic{equation}}
\section*{Appendix B: Two mode Squeezed State}
In this section, we employ the approach of Seifoory \textit{et al.} \cite{hossein} to show that the solution of the LME of \cref{master} for the Hamiltonian given in \cref{EqHprime} with $H_I^{red} = 0$ is a two mode squeezed thermal state. We will use two arbitrary modes with thermal populations denoted as $n_1$ and $n_2$, which correspond to the cavity and mechanical phonon thermal populations in the main text. We consider the generic form of the SPDC Hamiltonian
\begin{gather}
H = \hbar\omega_1 b_1^\dag b_1 + \hbar\omega_2 b_2^\dag b_2 + \hbar(\gamma b_1^\dag b_2^\dag + \gamma^* b_1 b_2)
\end{gather}
To see that the density operator that solves the LME is a two-mode STS, we first modify \cref{STS} to the form 
\begin{gather}
\rho(t) = S(\xi) \rho_T^{1/2}(n_1, n_2) O(t) \rho_T^{1/2}(n_1, n_2) S^\dag(\xi)
\end{gather}
and invert to obtain 
\begin{gather}
O(t) = \rho_T^{-1/2}(n_1, n_2)S^\dag(\xi)\rho(t) S(\xi)\rho_T^{-1/2}(n_1, n_2).
\end{gather}
For the solution to hold, the operator $O(t)$ must be the identity for all time. Using the product rule, the operator's time derivative can be split into five parts:
\begin{gather}
\dot{O}(t) = \dot{O}_T(t) + \dot{O}_S(t) + \dot{O}_0 + \dot{O}_V + \dot{O}_L.
\label{odot}
\end{gather}
The first term is defined using the thermal state change
\begin{gather}
\dot{O}_T = \frac{d\rho_T^{-1/2}}{dt} S^{\dag}\rho S \rho_T^{-1/2} + \rho_T^{-1/2} S^\dag \rho S \frac{d \rho_T^{-1/2}}{dt}
\end{gather}
or
\begin{gather}
\dot{O}_T = \acomm{J}{O}, 
\end{gather}
where
\begin{gather}
J \equiv \frac{d\rho_T^{-1/2}}{dt}\rho_T^{1/2} = \rho_T^{1/2}\frac{d\rho_T^{-1/2}}{dt}.
\end{gather}
For convenience, let $x_i = e^{-\beta\hbar\omega_i}$, which gives
\begin{gather}
J = \frac{1}{2}\sum_{i=1}^2 \frac{1}{x_i}\frac{dx_i}{dt}(n_i - b_i^\dag b_i).
\end{gather}

The second term contains the time evolution of the squeeze parameter
\begin{align}
\begin{aligned}
\dot{O}_S(t) = \rho_T^{-1/2} \frac{dS^\dag}{dt} S \rho_T^{1/2} (\rho_T^{-1/2} S^\dag \rho S \rho_T^{-1/2}) \\+ (\rho_T^{-1/2} S^\dag \rho S \rho_T^{-1/2})\rho_T^{1/2} S^\dag \frac{dS}{dt}\rho_T^{-1/2},
\end{aligned}
\end{align}
which can be rewritten as
\begin{gather}
\dot{O}_S(t) = LO + OL^\dag, 
\end{gather}
where
\begin{align}
\begin{aligned}
L \equiv &\rho_T^{-1/2}\frac{dS^\dag}{dt} S \rho_T^{1/2} \\
= \frac{1}{2}\big[ &\dot{u}(x_1^{-1/2}x_2^{-1/2}b_1^\dag b_2^\dag e^{i\phi_S} - x_1^{1/2}x_2^{1/2}b_1 b_2 e^{-i\phi_S}) \\ &+ ics\dot{\phi_S}(x_1^{-1/2}x_2^{-1/2}b_1^\dag b_2^\dag e^{i\phi_S} + x_1^{1/2}x_2^{1/2}b_1 b_2 e^{-i\phi_S}) \\ &- is^2 \dot{\phi_S}(1 + b_1^\dag b_1 + b_2^\dag b_2)\big]  % This is NOT trivial, and I AM fuzzy on the details. If there's major problems look here but I THINK this is equivilient to colin's paper, just with 2u -> u from defn of squeeze op.
\end{aligned}
\end{align}
and we have defined $s\equiv\sinh(u)$ and $c\equiv\cosh(u)$. Splitting this into commuting and anti-commuting parts $L = M + iN$, the term can be written as
\begin{gather}
\dot{O}_S = \acomm{M}{O} + i\comm{N}{O}, 
\end{gather}
where
\begin{align}
\begin{aligned}
M \equiv \frac{1}{4}&(x_1^{-1/2}x_2^{-1/2} - x_1^{1/2}x_2^{1/2}) \\ &\times\Big( b_1^\dag b_2^\dag(\dot{u} + ics\dot{\phi_S})e^{i\phi_S}  + b_1 b_2(\dot{u} - ics\dot{\phi_S})e^{-i\phi_S}\Big),
\end{aligned}
\end{align}
and we don't give the expression for $N$ as it will not contribute to the final result. The final three terms come from the derivative of the density operator, according to \cref{master}. The unperturbed Hamiltonian contributes the term
\begin{gather}
\dot{O}_0 = \rho_T^{-1/2}S^\dag \left( -i\comm{\omega_1 b_1^\dag b_1 + \omega_2 b_2^\dag b_2}{\rho} \right) S \rho_T^{-1/2},
\end{gather}
which can be rewritten as
\begin{gather}
\dot{O}_0 = -i \left(\omega_1 G_1 O - \omega_1 OG_1^\dag + \omega_2 G_2 O - \omega_2 OG_2^\dag \right), 
\end{gather}
with
\begin{gather}
G_i \equiv \rho_T^{-1/2}S^\dag b_i^\dag b_i S \rho_T^{1/2}.
\end{gather} 
Again, this can be split into commuting and anti-commuting parts as $G_i = P_i + iQ_i$, to obtain
\begin{gather}
\dot{O}_0 = -i\comm{\omega_1 P_1 + \omega_2 P_2}{O} + \acomm{\omega_1 Q_1 + \omega_2 Q_2}{O}.
\end{gather}
The perturbation, written as $V(t) =\hbar( \gamma b_1^\dag b_2^\dag + \gamma^* b_1 b_2)$, contributes
\begin{gather}
\dot{O}_V = \rho_T^{-1/2}S^\dag \left( -\frac{i}{\hbar}\comm{V}{\rho} \right) S \rho_T^{-1/2}
\end{gather}
and can be rewritten as

\begin{align}
&\dot{O}_V = -i\comm{\bar{P}}{O} + \acomm{\bar{Q}}{O},
\end{align} 
where
\begin{align}
&\begin{aligned}
\bar{P} \equiv &-cs(\gamma e^{-i\phi_S} + \gamma^*e^{i\phi_S})(1 + b_1^\dag b_1 + b_2^\dag) \\ &+ \frac{1}{2}(\gamma c^2 + \gamma^* s^2e^{2i\phi_S})(x_1^{-1/2}x_2^{-1/2} + x_1^{1/2}x_2^{1/2}){b^\dag}^2 \\ &+ \frac{1}{2}(\gamma s^2e^{-2i\phi_S} + \gamma^* c^2)(x_1^{-1/2}x_2^{-1/2} + x_1^{1/2}x_2^{1/2}) b^2, \end{aligned} 
\end{align}
\begin{align}
&\begin{aligned}
\bar{Q} \equiv &-\frac{i}{2}(x_1^{-1/2}x_2^{-1/2} - x_1^{1/2}x_2^{1/2})(\gamma^*s^2 e^{2i\phi_S} + \gamma c^2)b_1^\dag b_2^\dag \\ &+ \frac{i}{2}(x_1^{-1/2}x_2^{-1/2} - x_1^{1/2}x_2^{1/2})(\gamma s^2 e^{-2i\phi_S} + \gamma^* c^2)b_1 b_2.
\end{aligned}
\end{align} % Also NOT confident here, the terms are not great
Finally, the Lindblad terms contribute the term
\begin{widetext}
\begin{align}
\dot{O}_{L} = \sum_{i=1}^2 \left[
\begin{aligned}
\Gamma(n_i^b + 1) \Big( &\rho_T^{-1/2}S^\dag b_i \rho b_i^\dag S \rho_T^{-1/2} - \frac{1}{2}\rho_T^{-1/2}S^\dag (\acomm{b_i^\dag b_i}{\rho}) S \rho_T^{-1/2}\Big) \\
+ \Gamma n_i^b \Big( &\rho_T^{-1/2}S^\dag b_i^\dag \rho b_i S \rho_T^{-1/2} - \frac{1}{2}\rho_T^{-1/2}S^\dag (\acomm{b_i b_i^\dag}{\rho}) S \rho_T^{-1/2}\Big)
\end{aligned} \right].
\end{align}
\end{widetext}
Now, we define the operators
\begin{align}
& T_i \equiv \rho_T^{-1/2}S^\dag b_i S \rho_T^{1/2}, \\
& \tilde{T}_i \equiv \rho_T^{-1/2}S^\dag b_i^\dag S \rho_T^{1/2}, \\
& \begin{aligned}
\tilde{G}_i &\equiv \rho_T^{-1/2} S^\dag b_i b_i^\dag S \rho^{1/2} \\ &= G_i + \rho_T^{-1/2} S^\dag \comm{b_i}{b_i^\dag} S \rho_T^{1/2} \\ &= G_i + 1
\end{aligned}
\end{align}
and make the simplifications
\begin{gather}
\rho_T^{-1/2}S^\dag b_i \rho b_i^\dag S \rho_T^{-1/2} = T_i O T_i^\dag, \\
\rho_T^{-1/2}S^\dag b_i^\dag \rho b_i S \rho_T^{-1/2} = \tilde{T}_i O {\tilde{T}_i}^\dag,
\end{gather}
To obtain
\begin{align}
\dot{O}_{L} &= \sum_{i=1}^2 \left[ \begin{aligned}
\Gamma(n_i^b + 1) &\left[T_i O T_i^\dag - \frac{1}{2}(G_i O + OG_i^\dag)\right] \\
+ \Gamma n_i^b &\left[\tilde{T}_i O \tilde{T}_i^\dag -\frac{1}{2}(\tilde{G}_i O + O\tilde{G}_i^\dag)\right] 
\end{aligned} \right] \nonumber \\
% & = \sum_{i=1}^2 \left[ 
% \begin{aligned}
% \Gamma(n_i^b + 1) &\left[T_i O T_i^\dag - \frac{1}{2}(G_iO + OG_i^\dag)\right] \\ + \Gamma n_i^b &\left[\tilde{T}_i O \tilde{T}_i^\dag -\frac{1}{2}(G_i O + OG_i^\dag) - O\right] 
% \end{aligned} \right] \nonumber \\
& = \sum_{i=1}^2 \left[
\begin{aligned}
\Gamma(n_i^b + 1) \left[T_i O T_i^\dag - \frac{1}{2}\acomm{P_i}{O} - \frac{i}{2}\comm{Q_i}{O}\right]& \\+ \Gamma n_i^b \left[\tilde{T}_i O \tilde{T}_i^\dag -\frac{1}{2}\acomm{P_i}{O} - \frac{i}{2}\comm{Q_i}{O} - O\right]& 
\end{aligned} \right].
\end{align}
Now, inserting the parts of the equation back into \cref{odot}, then setting $O$ to the identity for all time yields
\begin{align}
\begin{aligned}
0 &= 2J + 2M + 2\omega_1 Q_1 + 2\omega_2 Q_2 + \frac{2}{\hbar}\bar{Q}  \\ &+ \sum_{i=1}^2 \left[\Gamma_i (n_i^b + 1)(T_iT_i^\dag - P_i) + \Gamma_i n_i^b (\tilde{T}_i \tilde{T}_i^\dag - P_i - 1) \right]. \label{simpleO}
\end{aligned}
\end{align}

Next, we introduce the Hermitian operators 
\begin{gather}
\chi_1 \equiv b_1^\dag b_2^\dag e^{i\phi_S} + b_1 b_2 e^{-i\phi_S}, \\
\chi_2 \equiv i(b_1^\dag b_2^\dag e^{i\phi_S} - b_1 b_2 e^{-i\phi_S}), 
\end{gather} 
and obtain the following expressions:
\begin{gather*}
T_1 T_1^\dag = x_1 c^2 b_1 b_1^\dag + x_2^{-1}s^2 b_2^\dag b_2 - x_1^{1/2}x_2^{-1/2}cs\chi_1, \\
T_2 T_2^\dag = x_1^{-1} s^2 b_1^\dag b_1 + x_2 c^2 b_2 b_2^\dag - x_1^{-1/2}x_2^{1/2}cs\chi_1, \\
\tilde{T}_1 \tilde{T}_1^\dag = x^{-1}_1 c^2 b^\dag_1 b_1 + x_2 s^2 b_2 b^\dag_2 - x_1^{-1/2}x_2^{1/2} cs \chi_1, \\
\tilde{T}_2 \tilde{T}_2^\dag = x_1 s^2 b_1 b^\dag_1 + x_2^{-1} c^2 b^\dag_2 b_2 - x_1^{1/2}x_2^{-1/2} cs \chi_1, \\
M = \frac{1}{4}(x_1^{-1/2}x_2^{-1/2} - x_1^{1/2}x_2^{1/2})(\dot{u}\chi_1 + cs\dot{\phi_S} \chi_2), \\
Q_1 = Q_2 = \frac{cs}{2}(x_1^{-1/2}x_2^{-1/2} - x_1^{1/2}x_2^{1/2}) \chi_2, 
\end{gather*}
\begin{align*}
&\begin{aligned} 2P_1 = s^2 &+ b_1^\dag b_1 c^2 + b_2^\dag b_2 s^2 \\ &- (x_1^{-1/2}x_2^{-1/2} + x_1^{1/2}x_2^{1/2})cs \chi_1, \\
2P_2 = s^2 &+ b_2^\dag b_2 c^2 + b_1^\dag b_1 s^2 \\ &- (x_1^{-1/2}x_2^{-1/2} + x_1^{1/2}x_2^{1/2})cs \chi_1, \end{aligned} \\
&\bar{Q} = \frac{-i}{4}(x_1^{-1/2}x_2^{-1/2} - x_1^{1/2}x_2^{1/2}) \\ &\times \left[
\begin{aligned} &(\gamma^*s^2e^{i\phi_S} + \gamma c^2e^{-i\phi_S})(\chi_1 - i\chi_2) \\ - &(\gamma s^2e^{-i\phi_S} + \gamma^* c^2 e^{i\phi_S})(\chi_1 + i\chi_2) \end{aligned} \right].
\end{align*}
Finally all these terms may be substituted into \cref{simpleO}, 
% \begin{align*}
% 0 = &\sum_{i=1}^2 \frac{1}{x_i}\frac{dx_i}{dt}(n_i - b_i^\dag b_i) \\ % J
% + &\frac{1}{2}(x_1^{-1/2}x_2^{-1/2} - x_1^{1/2}x_2^{1/2})(\dot{u}\chi_1 + cs\dot{\phi_S} \chi_2) \\ %M
% + &\frac{cs}{2}(\omega_1 + \omega_2)(x_1^{-1/2}x_2^{-1/2} - x_1^{1/2}x_2^{1/2}) \chi_2 \\ % Q
% - &\frac{i}{2}(x_1^{-1/2}x_2^{-1/2} - x_1^{1/2}x_2^{1/2})    \\ &\times \left[ \begin{aligned}
% &(\gamma^*s^2e^{i\phi_S} + \gamma c^2e^{-i\phi_S})(\chi_1 - i\chi_2) \\ - &(\gamma s^2e^{-i\phi_S} + ^*\gamma^* c^2 e^{i\phi_S})(\chi_1 + i\chi_2) \end{aligned} \right] \\ % Qbar
% + &\Gamma_1(n_1^b + 1) \Big[ x_1 c^2 b_1 b_1^\dag + x_2^{-1}s^2 b_2^\dag b_2 - x_1^{1/2}x_2^{-1/2}cs\chi_1 \Big] \\ % lind1 
% + & \Gamma_1 n_1^b \Big[ x^{-1}_1 c^2 b^\dag_1 b_1 + x_2 s^2 b_2 b^\dag_2 - x_1^{-1/2}x_2^{1/2} cs \chi_1  - 1\Big] \\
% + & \Gamma_2(n_2^b + 1) \Big[ x_1^{-1} s^2 b_1^\dag b_1 + x_2 c^2 b_2 b_2^\dag - x_1^{-1/2}x_2^{1/2}cs\chi_1 \Big] \\
% + & \Gamma_2 n_2^b \Big[ x_1 s^2 b_1 b^\dag_1 + x_2^{-1} c^2 b^\dag_2 b_2 - x_1^{1/2}x_2^{-1/2} cs \chi_1 - 1\Big] \\
% -\big(&\Gamma_1(2n_1^b + 1) + \Gamma_2(2n_2^b + 1)\big) \\ & \times \left[ \begin{aligned} s^2 &+ b_1^\dag b_1 c^2 + b_2^\dag b_2 s^2 \\ &- (x_1^{-1/2}x_2^{-1/2} + x_1^{1/2}x_2^{1/2})cs \chi_1) \end{aligned} \right]. % P Terms 
% \end{align*}
which allows us to form independent equations in $\chi_1$:
\begin{align}
\begin{aligned}
&0 = \frac{1}{2}(x_1^{-1/2}x_2^{-1/2} - x_1^{1/2}x_2^{1/2})\big(\dot{u} + i\gamma^* e^{i\phi_S} - i\gamma e^{-i\phi_S}\big) \\ &+ \frac{1}{2}cs(x_1^{-1/2}x_2^{-1/2} + x_1^{1/2}x_2^{1/2})\big(\Gamma_1(2n_1^b + 1) + \Gamma_2(2n_2^b + 1) \big) \\
&-cs\big( \Gamma_1 x_1^{1/2}x_2^{-1/2} + \Gamma_2 x_1^{-1/2}x_2^{1/2}\big) \\ &+ cs(x_1^{1/2}x_2^{-1/2} + x_1^{-1/2}x_2^{1/2})(\Gamma_1 n_1^b + \Gamma_1 n_2^b);
\end{aligned}
\end{align}
in $\chi_2$:
\begin{align}
\begin{aligned}
&0 = \frac{1}{2}(x_1^{-1/2}x_2^{-1/2} - x_1^{1/2}x_2^{1/2}) \left[ cs\dot{\phi_S} + cs(\omega_1 + \omega_2) \right] \\
&- \frac{1}{2}(x_1^{-1/2}x_2^{-1/2} - x_1^{1/2}x_2^{1/2}) (c^2 + s^2) (\gamma e^{-i\phi_S} + \gamma e^{i\phi_S});
\end{aligned}
\end{align}
in $b_1^\dag b_1$ and $b_2^\dag b_2$:
\begin{align}
\begin{aligned}
&0 = -x_1^{-1} \frac{dx_1}{dt} + (x_1 + x_1^{-1})(\Gamma_1 n_1^b c^2 + \Gamma_2 n_2^b s^2) \\
&- \big(\Gamma_1 (2n_1^b + 1)c^2 + \Gamma_2 (2n_2^b + 1)s^2\big) + x_1(\Gamma_1 c^2 + \Gamma_2 s^2);
\end{aligned} \\
\begin{aligned}
&0 = -x_2^{-1} \frac{dx_2}{dt} + (x_2 + x_2^{-1})(\Gamma_1 n_1^b s^2 + \Gamma_2 n_2^b c^2) \\
&- \big(\Gamma_1 (2n_1^b + 1) s^2 + \Gamma_2 (2n_2^b + 1)c^2\big) + x_2(\Gamma_1 s^2 + \Gamma_2 c^2);
\end{aligned}
\end{align}
and in the constant terms:
\begin{align}
\begin{aligned}
0 = &x_1n_1\frac{dx_1}{dt} + x_2n_2\frac{dx_2}{dt} \\ 
&- \big( \Gamma_1(2n_1^b + 1) + \Gamma_2(2n_2^b + 1)\big)s^2 \\ &+ \Gamma_1 n_1^b(x_1 c^2 + x_2 s^2) + \Gamma_2 n_2^b (x_1 s^2 + x_2 c^2) \\ &+ c^2(\Gamma_1 x_1 + \Gamma_2 x_2) + (\Gamma_1n_1^b + \Gamma_2 n_2^b).
\end{aligned}
\end{align}

These lead to the following equations of motion:
% \begin{align}
% \begin{aligned}
% \frac{du}{dt} = -&cs\frac{x_1^{-1/2}x_2^{-1/2} + x_1^{1/2}x_2^{1/2}}{x_1^{-1/2}x_2^{-1/2} - x_1^{1/2}x_2^{1/2}} \\ &\times \left( \Gamma_1(2n_1^b + 1) + \Gamma_2 (2n_2^b + 1)\right) \\ 
% + &2\frac{cs}{x_1^{-1/2}x_2^{-1/2} - x_1^{1/2}x_2^{1/2}} \\ &\times (\Gamma_1 x_1^{1/2}x_2^{-1/2} + \Gamma_2 x_1^{-1/2}x_2^{1/2} ) \\
% + &2cs\frac{x_1^{1/2}x_2^{-1/2} + x_1^{-1/2}x_2^{1/2}}{x_1^{-1/2}x_2^{-1/2} - x_1^{1/2}x_2^{1/2}} (\Gamma_1 n_1^b + \Gamma_2 n_2^b) \\
% + &i(\alpha\gamma e^{-i\phi} - \alpha^*\gamma^* e^{i\phi}) \\
% \end{aligned}
% \end{align}
\begin{align}
\begin{aligned}
\frac{du}{dt} = &i(\gamma e^{-i\phi_S} - \gamma^* e^{i\phi_S}) - cs\frac{n_2 - n_1}{n_1 + n_2 + 1}(\Gamma_1 - \Gamma_2) \\ &-cs\frac{ \Gamma_1 (2n_1^b + 1) + \Gamma_2 (2n_2^b + 1) }{n_1 + n_2 + 1},
\end{aligned}
\end{align}
\begin{gather}
\frac{d \phi_S}{dt} = \frac{c^2 + s^2}{cs}(\gamma e^{-i\phi_S} + \gamma^* e^{i\phi_S}) - \omega_1 - \omega_2,
\end{gather}
\begin{gather}
\frac{dn_1}{dt} = \Gamma_1 c^2 (n_1^b - n_1) + \Gamma_2 s^2(n_2^b + n_1 + 1), \\
\frac{dn_2}{dt} = \Gamma_2 c^2 (n_2^b - n_2) + \Gamma_1 s^2(n_1^b + n_2 + 1),
\end{gather}
Which are the dynamic equations given in Sec. \ref{squeezing}.

\bibliography{refs.bib}

\begin{thebibliography}{40}%
\makeatletter
\providecommand \@ifxundefined [1]{%
 \@ifx{#1\undefined}
}%
\providecommand \@ifnum [1]{%
 \ifnum #1\expandafter \@firstoftwo
 \else \expandafter \@secondoftwo
 \fi
}%
\providecommand \@ifx [1]{%
 \ifx #1\expandafter \@firstoftwo
 \else \expandafter \@secondoftwo
 \fi
}%
\providecommand \natexlab [1]{#1}%
\providecommand \enquote  [1]{``#1''}%
\providecommand \bibnamefont  [1]{#1}%
\providecommand \bibfnamefont [1]{#1}%
\providecommand \citenamefont [1]{#1}%
\providecommand \href@noop [0]{\@secondoftwo}%
\providecommand \href [0]{\begingroup \@sanitize@url \@href}%
\providecommand \@href[1]{\@@startlink{#1}\@@href}%
\providecommand \@@href[1]{\endgroup#1\@@endlink}%
\providecommand \@sanitize@url [0]{\catcode `\\12\catcode `\$12\catcode `\&12\catcode `\#12\catcode `\^12\catcode `\_12\catcode `\%12\relax}%
\providecommand \@@startlink[1]{}%
\providecommand \@@endlink[0]{}%
\providecommand \url  [0]{\begingroup\@sanitize@url \@url }%
\providecommand \@url [1]{\endgroup\@href {#1}{\urlprefix }}%
\providecommand \urlprefix  [0]{URL }%
\providecommand \Eprint [0]{\href }%
\providecommand \doibase [0]{https://doi.org/}%
\providecommand \selectlanguage [0]{\@gobble}%
\providecommand \bibinfo  [0]{\@secondoftwo}%
\providecommand \bibfield  [0]{\@secondoftwo}%
\providecommand \translation [1]{[#1]}%
\providecommand \BibitemOpen [0]{}%
\providecommand \bibitemStop [0]{}%
\providecommand \bibitemNoStop [0]{.\EOS\space}%
\providecommand \EOS [0]{\spacefactor3000\relax}%
\providecommand \BibitemShut  [1]{\csname bibitem#1\endcsname}%
\let\auto@bib@innerbib\@empty
%</preamble>
\bibitem [{\citenamefont {Zhang}\ \emph {et~al.}(2024)\citenamefont {Zhang}, \citenamefont {Wang}, \citenamefont {Zhang}, \citenamefont {Jiao}, \citenamefont {Zuo}, \citenamefont {\c{S}ahin K.~\"{O}zdemir}, \citenamefont {Qiu}, \citenamefont {Nori},\ and\ \citenamefont {Jing}}]{Zhang:24}%
  \BibitemOpen
  \bibfield  {author} {\bibinfo {author} {\bibfnamefont {S.-D.}\ \bibnamefont {Zhang}}, \bibinfo {author} {\bibfnamefont {J.}~\bibnamefont {Wang}}, \bibinfo {author} {\bibfnamefont {Q.}~\bibnamefont {Zhang}}, \bibinfo {author} {\bibfnamefont {Y.-F.}\ \bibnamefont {Jiao}}, \bibinfo {author} {\bibfnamefont {Y.-L.}\ \bibnamefont {Zuo}}, \bibinfo {author} {\bibnamefont {\c{S}ahin K.~\"{O}zdemir}}, \bibinfo {author} {\bibfnamefont {C.-W.}\ \bibnamefont {Qiu}}, \bibinfo {author} {\bibfnamefont {F.}~\bibnamefont {Nori}},\ and\ \bibinfo {author} {\bibfnamefont {H.}~\bibnamefont {Jing}},\ }\bibfield  {title} {\bibinfo {title} {Squeezing-enhanced quantum sensing with quadratic optomechanics},\ }\href {https://doi.org/10.1364/OPTICAQ.523480} {\bibfield  {journal} {\bibinfo  {journal} {Optica Quantum}\ }\textbf {\bibinfo {volume} {2}},\ \bibinfo {pages} {222} (\bibinfo {year} {2024})}\BibitemShut {NoStop}%
\bibitem [{\citenamefont {A.}\ \emph {et~al.}(2023)\citenamefont {A.}, \citenamefont {Kono}, \citenamefont {Chegnizadeh},\ and\ \citenamefont {Kippenberg}}]{mechOsc}%
  \BibitemOpen
  \bibfield  {author} {\bibinfo {author} {\bibfnamefont {Y.}~\bibnamefont {A.}}, \bibinfo {author} {\bibfnamefont {S.}~\bibnamefont {Kono}}, \bibinfo {author} {\bibfnamefont {M.}~\bibnamefont {Chegnizadeh}},\ and\ \bibinfo {author} {\bibfnamefont {T.~J.}\ \bibnamefont {Kippenberg}},\ }\bibfield  {title} {\bibinfo {title} {A squeezed mechanical oscillator with millisecond quantum decoherence},\ }\href {https://doi.org/10.1038/s41567-023-02135-y} {\bibfield  {journal} {\bibinfo  {journal} {Nat. Phys.}\ }\textbf {\bibinfo {volume} {19}},\ \bibinfo {pages} {1697} (\bibinfo {year} {2023})}\BibitemShut {NoStop}%
\bibitem [{\citenamefont {Dobrindt}\ \emph {et~al.}(2008)\citenamefont {Dobrindt}, \citenamefont {Wilson-Rae},\ and\ \citenamefont {Kippenberg}}]{PhysRevLett.101.263602}%
  \BibitemOpen
  \bibfield  {author} {\bibinfo {author} {\bibfnamefont {J.~M.}\ \bibnamefont {Dobrindt}}, \bibinfo {author} {\bibfnamefont {I.}~\bibnamefont {Wilson-Rae}},\ and\ \bibinfo {author} {\bibfnamefont {T.~J.}\ \bibnamefont {Kippenberg}},\ }\bibfield  {title} {\bibinfo {title} {Parametric normal-mode splitting in cavity optomechanics},\ }\href {https://doi.org/10.1103/PhysRevLett.101.263602} {\bibfield  {journal} {\bibinfo  {journal} {Phys. Rev. Lett.}\ }\textbf {\bibinfo {volume} {101}},\ \bibinfo {pages} {263602} (\bibinfo {year} {2008})}\BibitemShut {NoStop}%
\bibitem [{\citenamefont {Nunnenkamp}\ \emph {et~al.}(2012)\citenamefont {Nunnenkamp}, \citenamefont {B\o{}rkje},\ and\ \citenamefont {Girvin}}]{PhysRevA.85.051803}%
  \BibitemOpen
  \bibfield  {author} {\bibinfo {author} {\bibfnamefont {A.}~\bibnamefont {Nunnenkamp}}, \bibinfo {author} {\bibfnamefont {K.}~\bibnamefont {B\o{}rkje}},\ and\ \bibinfo {author} {\bibfnamefont {S.~M.}\ \bibnamefont {Girvin}},\ }\bibfield  {title} {\bibinfo {title} {Cooling in the single-photon strong-coupling regime of cavity optomechanics},\ }\href {https://doi.org/10.1103/PhysRevA.85.051803} {\bibfield  {journal} {\bibinfo  {journal} {Phys. Rev. A}\ }\textbf {\bibinfo {volume} {85}},\ \bibinfo {pages} {051803} (\bibinfo {year} {2012})}\BibitemShut {NoStop}%
\bibitem [{\citenamefont {Marquardt}\ \emph {et~al.}(2007)\citenamefont {Marquardt}, \citenamefont {Chen}, \citenamefont {Clerk},\ and\ \citenamefont {Girvin}}]{PhysRevLett.99.093902}%
  \BibitemOpen
  \bibfield  {author} {\bibinfo {author} {\bibfnamefont {F.}~\bibnamefont {Marquardt}}, \bibinfo {author} {\bibfnamefont {J.~P.}\ \bibnamefont {Chen}}, \bibinfo {author} {\bibfnamefont {A.~A.}\ \bibnamefont {Clerk}},\ and\ \bibinfo {author} {\bibfnamefont {S.~M.}\ \bibnamefont {Girvin}},\ }\bibfield  {title} {\bibinfo {title} {Quantum theory of cavity-assisted sideband cooling of mechanical motion},\ }\href {https://doi.org/10.1103/PhysRevLett.99.093902} {\bibfield  {journal} {\bibinfo  {journal} {Phys. Rev. Lett.}\ }\textbf {\bibinfo {volume} {99}},\ \bibinfo {pages} {093902} (\bibinfo {year} {2007})}\BibitemShut {NoStop}%
\bibitem [{\citenamefont {Kronwald}\ \emph {et~al.}(2013)\citenamefont {Kronwald}, \citenamefont {Marquardt},\ and\ \citenamefont {Clerk}}]{PhysRevA.88.063833}%
  \BibitemOpen
  \bibfield  {author} {\bibinfo {author} {\bibfnamefont {A.}~\bibnamefont {Kronwald}}, \bibinfo {author} {\bibfnamefont {F.}~\bibnamefont {Marquardt}},\ and\ \bibinfo {author} {\bibfnamefont {A.~A.}\ \bibnamefont {Clerk}},\ }\bibfield  {title} {\bibinfo {title} {Arbitrarily large steady-state bosonic squeezing via dissipation},\ }\href {https://doi.org/10.1103/PhysRevA.88.063833} {\bibfield  {journal} {\bibinfo  {journal} {Phys. Rev. A}\ }\textbf {\bibinfo {volume} {88}},\ \bibinfo {pages} {063833} (\bibinfo {year} {2013})}\BibitemShut {NoStop}%
\bibitem [{\citenamefont {Basilewitsch}\ \emph {et~al.}(2019)\citenamefont {Basilewitsch}, \citenamefont {Koch},\ and\ \citenamefont {Reich}}]{https://doi.org/10.1002/qute.201800110}%
  \BibitemOpen
  \bibfield  {author} {\bibinfo {author} {\bibfnamefont {D.}~\bibnamefont {Basilewitsch}}, \bibinfo {author} {\bibfnamefont {C.~P.}\ \bibnamefont {Koch}},\ and\ \bibinfo {author} {\bibfnamefont {D.~M.}\ \bibnamefont {Reich}},\ }\bibfield  {title} {\bibinfo {title} {Quantum optimal control for mixed state squeezing in cavity optomechanics},\ }\href {https://doi.org/https://doi.org/10.1002/qute.201800110} {\bibfield  {journal} {\bibinfo  {journal} {Advanced Quantum Technologies}\ }\textbf {\bibinfo {volume} {2}},\ \bibinfo {pages} {1800110} (\bibinfo {year} {2019})}\BibitemShut {NoStop}%
\bibitem [{\citenamefont {Halaski}\ \emph {et~al.}(2024)\citenamefont {Halaski}, \citenamefont {Krauss}, \citenamefont {Basilewitsch},\ and\ \citenamefont {Koch}}]{PhysRevA.110.013512}%
  \BibitemOpen
  \bibfield  {author} {\bibinfo {author} {\bibfnamefont {A.}~\bibnamefont {Halaski}}, \bibinfo {author} {\bibfnamefont {M.~G.}\ \bibnamefont {Krauss}}, \bibinfo {author} {\bibfnamefont {D.}~\bibnamefont {Basilewitsch}},\ and\ \bibinfo {author} {\bibfnamefont {C.~P.}\ \bibnamefont {Koch}},\ }\bibfield  {title} {\bibinfo {title} {Quantum optimal control of squeezing in cavity optomechanics},\ }\href {https://doi.org/10.1103/PhysRevA.110.013512} {\bibfield  {journal} {\bibinfo  {journal} {Phys. Rev. A}\ }\textbf {\bibinfo {volume} {110}},\ \bibinfo {pages} {013512} (\bibinfo {year} {2024})}\BibitemShut {NoStop}%
\bibitem [{\citenamefont {Wang}\ and\ \citenamefont {Clerk}(2013)}]{PhysRevLett.110.253601}%
  \BibitemOpen
  \bibfield  {author} {\bibinfo {author} {\bibfnamefont {Y.-D.}\ \bibnamefont {Wang}}\ and\ \bibinfo {author} {\bibfnamefont {A.~A.}\ \bibnamefont {Clerk}},\ }\bibfield  {title} {\bibinfo {title} {Reservoir-engineered entanglement in optomechanical systems},\ }\href {https://doi.org/10.1103/PhysRevLett.110.253601} {\bibfield  {journal} {\bibinfo  {journal} {Phys. Rev. Lett.}\ }\textbf {\bibinfo {volume} {110}},\ \bibinfo {pages} {253601} (\bibinfo {year} {2013})}\BibitemShut {NoStop}%
\bibitem [{\citenamefont {Hofer}\ \emph {et~al.}(2011)\citenamefont {Hofer}, \citenamefont {Wieczorek}, \citenamefont {Aspelmeyer},\ and\ \citenamefont {Hammerer}}]{PhysRevA.84.052327}%
  \BibitemOpen
  \bibfield  {author} {\bibinfo {author} {\bibfnamefont {S.~G.}\ \bibnamefont {Hofer}}, \bibinfo {author} {\bibfnamefont {W.}~\bibnamefont {Wieczorek}}, \bibinfo {author} {\bibfnamefont {M.}~\bibnamefont {Aspelmeyer}},\ and\ \bibinfo {author} {\bibfnamefont {K.}~\bibnamefont {Hammerer}},\ }\bibfield  {title} {\bibinfo {title} {Quantum entanglement and teleportation in pulsed cavity optomechanics},\ }\href {https://doi.org/10.1103/PhysRevA.84.052327} {\bibfield  {journal} {\bibinfo  {journal} {Phys. Rev. A}\ }\textbf {\bibinfo {volume} {84}},\ \bibinfo {pages} {052327} (\bibinfo {year} {2011})}\BibitemShut {NoStop}%
\bibitem [{\citenamefont {Teufel}\ \emph {et~al.}(2011)\citenamefont {Teufel}, \citenamefont {Donner}, \citenamefont {Li}, \citenamefont {Harlow}, \citenamefont {Allman}, \citenamefont {Cicak}, \citenamefont {Sirois}, \citenamefont {Whittaker}, \citenamefont {Lehnert},\ and\ \citenamefont {Simmonds}}]{teufelCooling}%
  \BibitemOpen
  \bibfield  {author} {\bibinfo {author} {\bibfnamefont {J.~D.}\ \bibnamefont {Teufel}}, \bibinfo {author} {\bibfnamefont {T.}~\bibnamefont {Donner}}, \bibinfo {author} {\bibfnamefont {D.}~\bibnamefont {Li}}, \bibinfo {author} {\bibfnamefont {J.~W.}\ \bibnamefont {Harlow}}, \bibinfo {author} {\bibfnamefont {M.~S.}\ \bibnamefont {Allman}}, \bibinfo {author} {\bibfnamefont {K.}~\bibnamefont {Cicak}}, \bibinfo {author} {\bibfnamefont {A.~J.}\ \bibnamefont {Sirois}}, \bibinfo {author} {\bibfnamefont {J.~D.}\ \bibnamefont {Whittaker}}, \bibinfo {author} {\bibfnamefont {K.~W.}\ \bibnamefont {Lehnert}},\ and\ \bibinfo {author} {\bibfnamefont {R.~W.}\ \bibnamefont {Simmonds}},\ }\bibfield  {title} {\bibinfo {title} {Sideband cooling of micromechanical motion to the quantum ground state},\ }\href {https://doi.org/10.1038/nature10261} {\bibfield  {journal} {\bibinfo  {journal} {Nature}\ }\textbf {\bibinfo {volume} {475}},\ \bibinfo {pages} {359} (\bibinfo {year} {2011})}\BibitemShut {NoStop}%
\bibitem [{\citenamefont {Seis}\ \emph {et~al.}(2022)\citenamefont {Seis}, \citenamefont {Capelle}, \citenamefont {Langman}, \citenamefont {Saarinen}, \citenamefont {Planz},\ and\ \citenamefont {Schliesser}}]{electromechCooling}%
  \BibitemOpen
  \bibfield  {author} {\bibinfo {author} {\bibfnamefont {Y.}~\bibnamefont {Seis}}, \bibinfo {author} {\bibfnamefont {T.}~\bibnamefont {Capelle}}, \bibinfo {author} {\bibfnamefont {E.}~\bibnamefont {Langman}}, \bibinfo {author} {\bibfnamefont {S.}~\bibnamefont {Saarinen}}, \bibinfo {author} {\bibfnamefont {E.}~\bibnamefont {Planz}},\ and\ \bibinfo {author} {\bibfnamefont {A.}~\bibnamefont {Schliesser}},\ }\bibfield  {title} {\bibinfo {title} {Ground state cooling of an ultracoherent electromechanical system},\ }\href {https://doi.org/10.1038/s41467-022-29115-9} {\bibfield  {journal} {\bibinfo  {journal} {Nat. Commun.}\ }\textbf {\bibinfo {volume} {13}},\ \bibinfo {pages} {1507} (\bibinfo {year} {2022})}\BibitemShut {NoStop}%
\bibitem [{\citenamefont {Palomaki}\ \emph {et~al.}(2013{\natexlab{a}})\citenamefont {Palomaki}, \citenamefont {Teufel}, \citenamefont {Simmonds},\ and\ \citenamefont {Lehnert}}]{doi:10.1126/science.1244563}%
  \BibitemOpen
  \bibfield  {author} {\bibinfo {author} {\bibfnamefont {T.~A.}\ \bibnamefont {Palomaki}}, \bibinfo {author} {\bibfnamefont {J.~D.}\ \bibnamefont {Teufel}}, \bibinfo {author} {\bibfnamefont {R.~W.}\ \bibnamefont {Simmonds}},\ and\ \bibinfo {author} {\bibfnamefont {K.~W.}\ \bibnamefont {Lehnert}},\ }\bibfield  {title} {\bibinfo {title} {Entangling mechanical motion with microwave fields},\ }\href {https://doi.org/10.1126/science.1244563} {\bibfield  {journal} {\bibinfo  {journal} {Science}\ }\textbf {\bibinfo {volume} {342}},\ \bibinfo {pages} {710} (\bibinfo {year} {2013}{\natexlab{a}})}\BibitemShut {NoStop}%
\bibitem [{\citenamefont {Marti}\ \emph {et~al.}(2024)\citenamefont {Marti}, \citenamefont {Lupke}, \citenamefont {Joshi}, \citenamefont {Yang}, \citenamefont {Bild}, \citenamefont {Omahen}, \citenamefont {Chu},\ and\ \citenamefont {Fadel}}]{kerrMech}%
  \BibitemOpen
  \bibfield  {author} {\bibinfo {author} {\bibfnamefont {S.}~\bibnamefont {Marti}}, \bibinfo {author} {\bibfnamefont {U.}~\bibnamefont {Lupke}}, \bibinfo {author} {\bibfnamefont {O.}~\bibnamefont {Joshi}}, \bibinfo {author} {\bibfnamefont {Y.}~\bibnamefont {Yang}}, \bibinfo {author} {\bibfnamefont {M.}~\bibnamefont {Bild}}, \bibinfo {author} {\bibfnamefont {A.}~\bibnamefont {Omahen}}, \bibinfo {author} {\bibfnamefont {Y.}~\bibnamefont {Chu}},\ and\ \bibinfo {author} {\bibfnamefont {M.}~\bibnamefont {Fadel}},\ }\bibfield  {title} {\bibinfo {title} {Quantum squeezing in a nonlinear mechanical oscillator},\ }\href {https://doi.org/10.1038/s41567-024-02545-6} {\bibfield  {journal} {\bibinfo  {journal} {Nat. Phys.}\ }\textbf {\bibinfo {volume} {20}},\ \bibinfo {pages} {1448} (\bibinfo {year} {2024})}\BibitemShut {NoStop}%
\bibitem [{\citenamefont {Delaney}\ \emph {et~al.}(2019)\citenamefont {Delaney}, \citenamefont {Reed}, \citenamefont {Andrews},\ and\ \citenamefont {Lehnert}}]{PhysRevLett.123.183603}%
  \BibitemOpen
  \bibfield  {author} {\bibinfo {author} {\bibfnamefont {R.~D.}\ \bibnamefont {Delaney}}, \bibinfo {author} {\bibfnamefont {A.~P.}\ \bibnamefont {Reed}}, \bibinfo {author} {\bibfnamefont {R.~W.}\ \bibnamefont {Andrews}},\ and\ \bibinfo {author} {\bibfnamefont {K.~W.}\ \bibnamefont {Lehnert}},\ }\bibfield  {title} {\bibinfo {title} {Measurement of motion beyond the quantum limit by transient amplification},\ }\href {https://doi.org/10.1103/PhysRevLett.123.183603} {\bibfield  {journal} {\bibinfo  {journal} {Phys. Rev. Lett.}\ }\textbf {\bibinfo {volume} {123}},\ \bibinfo {pages} {183603} (\bibinfo {year} {2019})}\BibitemShut {NoStop}%
\bibitem [{\citenamefont {Xiang}\ \emph {et~al.}(2011)\citenamefont {Xiang}, \citenamefont {Song}, \citenamefont {Wen},\ and\ \citenamefont {Shi}}]{entanglementTMSTS}%
  \BibitemOpen
  \bibfield  {author} {\bibinfo {author} {\bibfnamefont {S.~H.}\ \bibnamefont {Xiang}}, \bibinfo {author} {\bibfnamefont {K.~H.}\ \bibnamefont {Song}}, \bibinfo {author} {\bibfnamefont {W.}~\bibnamefont {Wen}},\ and\ \bibinfo {author} {\bibfnamefont {Z.~G.}\ \bibnamefont {Shi}},\ }\bibfield  {title} {\bibinfo {title} {Entanglement behaviors of two-mode squeezed thermal states in two different environments},\ }\href {https://doi.org/10.1140/epjd/e2011-10546-1} {\bibfield  {journal} {\bibinfo  {journal} {Eur. Phys. J. D}\ }\textbf {\bibinfo {volume} {62}},\ \bibinfo {pages} {289} (\bibinfo {year} {2011})}\BibitemShut {NoStop}%
\bibitem [{\citenamefont {Prauzner-Bechcicki}(2004)}]{TMSVSres}%
  \BibitemOpen
  \bibfield  {author} {\bibinfo {author} {\bibfnamefont {J.~S.}\ \bibnamefont {Prauzner-Bechcicki}},\ }\bibfield  {title} {\bibinfo {title} {Two-mode squeezed vacuum state coupled to the common thermal reservoir},\ }\href {https://doi.org/10.1088/0305-4470/37/15/L04} {\bibfield  {journal} {\bibinfo  {journal} {Journal of Physics A: Mathematical and General}\ }\textbf {\bibinfo {volume} {37}},\ \bibinfo {pages} {L173} (\bibinfo {year} {2004})}\BibitemShut {NoStop}%
\bibitem [{\citenamefont {Xiang}\ \emph {et~al.}(2008)\citenamefont {Xiang}, \citenamefont {Shao},\ and\ \citenamefont {Song}}]{PhysRevA.78.052313}%
  \BibitemOpen
  \bibfield  {author} {\bibinfo {author} {\bibfnamefont {S.-H.}\ \bibnamefont {Xiang}}, \bibinfo {author} {\bibfnamefont {B.}~\bibnamefont {Shao}},\ and\ \bibinfo {author} {\bibfnamefont {K.-H.}\ \bibnamefont {Song}},\ }\bibfield  {title} {\bibinfo {title} {Environment-assisted creation and enhancement of two-mode entanglement in a joint decoherent model},\ }\href {https://doi.org/10.1103/PhysRevA.78.052313} {\bibfield  {journal} {\bibinfo  {journal} {Phys. Rev. A}\ }\textbf {\bibinfo {volume} {78}},\ \bibinfo {pages} {052313} (\bibinfo {year} {2008})}\BibitemShut {NoStop}%
\bibitem [{\citenamefont {Paz}\ and\ \citenamefont {Roncaglia}(2008)}]{PhysRevLett.100.220401}%
  \BibitemOpen
  \bibfield  {author} {\bibinfo {author} {\bibfnamefont {J.~P.}\ \bibnamefont {Paz}}\ and\ \bibinfo {author} {\bibfnamefont {A.~J.}\ \bibnamefont {Roncaglia}},\ }\bibfield  {title} {\bibinfo {title} {Dynamics of the entanglement between two oscillators in the same environment},\ }\href {https://doi.org/10.1103/PhysRevLett.100.220401} {\bibfield  {journal} {\bibinfo  {journal} {Phys. Rev. Lett.}\ }\textbf {\bibinfo {volume} {100}},\ \bibinfo {pages} {220401} (\bibinfo {year} {2008})}\BibitemShut {NoStop}%
\bibitem [{\citenamefont {Ayehu}(2024)}]{AYEHU2024100605}%
  \BibitemOpen
  \bibfield  {author} {\bibinfo {author} {\bibfnamefont {D.}~\bibnamefont {Ayehu}},\ }\bibfield  {title} {\bibinfo {title} {Optomechanical squeezing and entanglement in cavities optomechanical system},\ }\href {https://doi.org/https://doi.org/10.1016/j.rio.2024.100605} {\bibfield  {journal} {\bibinfo  {journal} {Results in Optics}\ }\textbf {\bibinfo {volume} {14}},\ \bibinfo {pages} {100605} (\bibinfo {year} {2024})}\BibitemShut {NoStop}%
\bibitem [{\citenamefont {Huang}\ and\ \citenamefont {Chen}(2018)}]{PhysRevA.98.063843}%
  \BibitemOpen
  \bibfield  {author} {\bibinfo {author} {\bibfnamefont {S.}~\bibnamefont {Huang}}\ and\ \bibinfo {author} {\bibfnamefont {A.}~\bibnamefont {Chen}},\ }\bibfield  {title} {\bibinfo {title} {Quadrature-squeezed light and optomechanical entanglement in a dissipative optomechanical system with a mechanical parametric drive},\ }\href {https://doi.org/10.1103/PhysRevA.98.063843} {\bibfield  {journal} {\bibinfo  {journal} {Phys. Rev. A}\ }\textbf {\bibinfo {volume} {98}},\ \bibinfo {pages} {063843} (\bibinfo {year} {2018})}\BibitemShut {NoStop}%
\bibitem [{\citenamefont {Nunnenkamp}\ \emph {et~al.}(2011)\citenamefont {Nunnenkamp}, \citenamefont {B\o{}rkje},\ and\ \citenamefont {Girvin}}]{PhysRevLett.107.063602}%
  \BibitemOpen
  \bibfield  {author} {\bibinfo {author} {\bibfnamefont {A.}~\bibnamefont {Nunnenkamp}}, \bibinfo {author} {\bibfnamefont {K.}~\bibnamefont {B\o{}rkje}},\ and\ \bibinfo {author} {\bibfnamefont {S.~M.}\ \bibnamefont {Girvin}},\ }\bibfield  {title} {\bibinfo {title} {Single-photon optomechanics},\ }\href {https://doi.org/10.1103/PhysRevLett.107.063602} {\bibfield  {journal} {\bibinfo  {journal} {Phys. Rev. Lett.}\ }\textbf {\bibinfo {volume} {107}},\ \bibinfo {pages} {063602} (\bibinfo {year} {2011})}\BibitemShut {NoStop}%
\bibitem [{\citenamefont {Johansson}\ \emph {et~al.}(2012)\citenamefont {Johansson}, \citenamefont {Nation},\ and\ \citenamefont {Nori}}]{JOHANSSON20121760}%
  \BibitemOpen
  \bibfield  {author} {\bibinfo {author} {\bibfnamefont {J.}~\bibnamefont {Johansson}}, \bibinfo {author} {\bibfnamefont {P.}~\bibnamefont {Nation}},\ and\ \bibinfo {author} {\bibfnamefont {F.}~\bibnamefont {Nori}},\ }\bibfield  {title} {\bibinfo {title} {Qutip: An open-source python framework for the dynamics of open quantum systems},\ }\href {https://doi.org/https://doi.org/10.1016/j.cpc.2012.02.021} {\bibfield  {journal} {\bibinfo  {journal} {Computer Physics Communications}\ }\textbf {\bibinfo {volume} {183}},\ \bibinfo {pages} {1760} (\bibinfo {year} {2012})}\BibitemShut {NoStop}%
\bibitem [{\citenamefont {Johansson}\ \emph {et~al.}(2013)\citenamefont {Johansson}, \citenamefont {Nation},\ and\ \citenamefont {Nori}}]{JOHANSSON20131234}%
  \BibitemOpen
  \bibfield  {author} {\bibinfo {author} {\bibfnamefont {J.}~\bibnamefont {Johansson}}, \bibinfo {author} {\bibfnamefont {P.}~\bibnamefont {Nation}},\ and\ \bibinfo {author} {\bibfnamefont {F.}~\bibnamefont {Nori}},\ }\bibfield  {title} {\bibinfo {title} {Qutip 2: A python framework for the dynamics of open quantum systems},\ }\href {https://doi.org/https://doi.org/10.1016/j.cpc.2012.11.019} {\bibfield  {journal} {\bibinfo  {journal} {Computer Physics Communications}\ }\textbf {\bibinfo {volume} {184}},\ \bibinfo {pages} {1234} (\bibinfo {year} {2013})}\BibitemShut {NoStop}%
\bibitem [{\citenamefont {Kr{\"a}mer}\ \emph {et~al.}(2018)\citenamefont {Kr{\"a}mer}, \citenamefont {Plankensteiner}, \citenamefont {Ostermann},\ and\ \citenamefont {Ritsch}}]{kramer2018quantumoptics}%
  \BibitemOpen
  \bibfield  {author} {\bibinfo {author} {\bibfnamefont {S.}~\bibnamefont {Kr{\"a}mer}}, \bibinfo {author} {\bibfnamefont {D.}~\bibnamefont {Plankensteiner}}, \bibinfo {author} {\bibfnamefont {L.}~\bibnamefont {Ostermann}},\ and\ \bibinfo {author} {\bibfnamefont {H.}~\bibnamefont {Ritsch}},\ }\bibfield  {title} {\bibinfo {title} {Quantumoptics. jl: A julia framework for simulating open quantum systems},\ }\href@noop {} {\bibfield  {journal} {\bibinfo  {journal} {Computer Physics Communications}\ }\textbf {\bibinfo {volume} {227}},\ \bibinfo {pages} {109} (\bibinfo {year} {2018})}\BibitemShut {NoStop}%
\bibitem [{\citenamefont {Wise}\ \emph {et~al.}(2024)\citenamefont {Wise}, \citenamefont {Dutreix},\ and\ \citenamefont {Pistolesi}}]{PhysRevA.109.L051501}%
  \BibitemOpen
  \bibfield  {author} {\bibinfo {author} {\bibfnamefont {J.~L.}\ \bibnamefont {Wise}}, \bibinfo {author} {\bibfnamefont {C.}~\bibnamefont {Dutreix}},\ and\ \bibinfo {author} {\bibfnamefont {F.}~\bibnamefont {Pistolesi}},\ }\bibfield  {title} {\bibinfo {title} {Nonclassical mechanical states in cavity optomechanics in the single-photon strong-coupling regime},\ }\href {https://doi.org/10.1103/PhysRevA.109.L051501} {\bibfield  {journal} {\bibinfo  {journal} {Phys. Rev. A}\ }\textbf {\bibinfo {volume} {109}},\ \bibinfo {pages} {L051501} (\bibinfo {year} {2024})}\BibitemShut {NoStop}%
\bibitem [{\citenamefont {Seifoory}\ \emph {et~al.}(2017)\citenamefont {Seifoory}, \citenamefont {Doutre}, \citenamefont {Dignam},\ and\ \citenamefont {Sipe}}]{hossein}%
  \BibitemOpen
  \bibfield  {author} {\bibinfo {author} {\bibfnamefont {H.}~\bibnamefont {Seifoory}}, \bibinfo {author} {\bibfnamefont {S.}~\bibnamefont {Doutre}}, \bibinfo {author} {\bibfnamefont {M.~M.}\ \bibnamefont {Dignam}},\ and\ \bibinfo {author} {\bibfnamefont {J.~E.}\ \bibnamefont {Sipe}},\ }\bibfield  {title} {\bibinfo {title} {Squeezed thermal states: the result of parametric down conversion in lossy cavities},\ }\href {https://doi.org/https://doi.org/10.1364/JOSAB.34.001587} {\bibfield  {journal} {\bibinfo  {journal} {Journal of the Optical Society of America B}\ }\textbf {\bibinfo {volume} {34}},\ \bibinfo {pages} {1587} (\bibinfo {year} {2017})}\BibitemShut {NoStop}%
\bibitem [{\citenamefont {Vendromin}\ and\ \citenamefont {Dignam}(2021)}]{PhysRevA.103.022418}%
  \BibitemOpen
  \bibfield  {author} {\bibinfo {author} {\bibfnamefont {C.}~\bibnamefont {Vendromin}}\ and\ \bibinfo {author} {\bibfnamefont {M.~M.}\ \bibnamefont {Dignam}},\ }\bibfield  {title} {\bibinfo {title} {Continuous-variable entanglement in a two-mode lossy cavity: An analytic solution},\ }\href {https://doi.org/10.1103/PhysRevA.103.022418} {\bibfield  {journal} {\bibinfo  {journal} {Phys. Rev. A}\ }\textbf {\bibinfo {volume} {103}},\ \bibinfo {pages} {022418} (\bibinfo {year} {2021})}\BibitemShut {NoStop}%
\bibitem [{\citenamefont {Aspelmeyer}\ \emph {et~al.}(2014)\citenamefont {Aspelmeyer}, \citenamefont {Kippenberg},\ and\ \citenamefont {Marquardt}}]{cavityOptomechanicsRMP}%
  \BibitemOpen
  \bibfield  {author} {\bibinfo {author} {\bibfnamefont {M.}~\bibnamefont {Aspelmeyer}}, \bibinfo {author} {\bibfnamefont {T.~J.}\ \bibnamefont {Kippenberg}},\ and\ \bibinfo {author} {\bibfnamefont {F.}~\bibnamefont {Marquardt}},\ }\bibfield  {title} {\bibinfo {title} {Cavity optomechanics},\ }\href {https://doi.org/10.1103/RevModPhys.86.1391} {\bibfield  {journal} {\bibinfo  {journal} {Rev. Mod. Phys.}\ }\textbf {\bibinfo {volume} {86}},\ \bibinfo {pages} {1391} (\bibinfo {year} {2014})}\BibitemShut {NoStop}%
\bibitem [{\citenamefont {Breuer}\ and\ \citenamefont {Petruccione}(2007)}]{OpenQuantum}%
  \BibitemOpen
  \bibfield  {author} {\bibinfo {author} {\bibfnamefont {H.-P.}\ \bibnamefont {Breuer}}\ and\ \bibinfo {author} {\bibfnamefont {F.}~\bibnamefont {Petruccione}},\ }\href {https://doi.org/10.1093/acprof:oso/9780199213900.001.0001} {\emph {\bibinfo {title} {{The Theory of Open Quantum Systems}}}}\ (\bibinfo  {publisher} {Oxford University Press},\ \bibinfo {year} {2007})\BibitemShut {NoStop}%
\bibitem [{\citenamefont {Hu}\ \emph {et~al.}(2015)\citenamefont {Hu}, \citenamefont {Huang}, \citenamefont {Liao}, \citenamefont {Tian},\ and\ \citenamefont {Goan}}]{PhysRevA.91.013812}%
  \BibitemOpen
  \bibfield  {author} {\bibinfo {author} {\bibfnamefont {D.}~\bibnamefont {Hu}}, \bibinfo {author} {\bibfnamefont {S.-Y.}\ \bibnamefont {Huang}}, \bibinfo {author} {\bibfnamefont {J.-Q.}\ \bibnamefont {Liao}}, \bibinfo {author} {\bibfnamefont {L.}~\bibnamefont {Tian}},\ and\ \bibinfo {author} {\bibfnamefont {H.-S.}\ \bibnamefont {Goan}},\ }\bibfield  {title} {\bibinfo {title} {Quantum coherence in ultrastrong optomechanics},\ }\href {https://doi.org/10.1103/PhysRevA.91.013812} {\bibfield  {journal} {\bibinfo  {journal} {Phys. Rev. A}\ }\textbf {\bibinfo {volume} {91}},\ \bibinfo {pages} {013812} (\bibinfo {year} {2015})}\BibitemShut {NoStop}%
\bibitem [{\citenamefont {Naseem}\ \emph {et~al.}(2018)\citenamefont {Naseem}, \citenamefont {Xuereb},\ and\ \citenamefont {M\"ustecapl\ifmmode \imath \else \i \fi{}o\ifmmode~\breve{g}\else \u{g}\fi{}lu}}]{PhysRevA.98.052123}%
  \BibitemOpen
  \bibfield  {author} {\bibinfo {author} {\bibfnamefont {M.~T.}\ \bibnamefont {Naseem}}, \bibinfo {author} {\bibfnamefont {A.}~\bibnamefont {Xuereb}},\ and\ \bibinfo {author} {\bibfnamefont {O.~E.}\ \bibnamefont {M\"ustecapl\ifmmode \imath \else \i \fi{}o\ifmmode~\breve{g}\else \u{g}\fi{}lu}},\ }\bibfield  {title} {\bibinfo {title} {Thermodynamic consistency of the optomechanical master equation},\ }\href {https://doi.org/10.1103/PhysRevA.98.052123} {\bibfield  {journal} {\bibinfo  {journal} {Phys. Rev. A}\ }\textbf {\bibinfo {volume} {98}},\ \bibinfo {pages} {052123} (\bibinfo {year} {2018})}\BibitemShut {NoStop}%
\bibitem [{\citenamefont {Palomaki}\ \emph {et~al.}(2013{\natexlab{b}})\citenamefont {Palomaki}, \citenamefont {Harlow}, \citenamefont {Teufel}, \citenamefont {Simmonds},\ and\ \citenamefont {Lehnert}}]{coherentStateTransfer}%
  \BibitemOpen
  \bibfield  {author} {\bibinfo {author} {\bibfnamefont {T.~A.}\ \bibnamefont {Palomaki}}, \bibinfo {author} {\bibfnamefont {J.~W.}\ \bibnamefont {Harlow}}, \bibinfo {author} {\bibfnamefont {J.~D.}\ \bibnamefont {Teufel}}, \bibinfo {author} {\bibfnamefont {R.~W.}\ \bibnamefont {Simmonds}},\ and\ \bibinfo {author} {\bibfnamefont {K.~W.}\ \bibnamefont {Lehnert}},\ }\bibfield  {title} {\bibinfo {title} {Coherent state transfer between itinerant microwave fields and a mechanical oscillator},\ }\href {https://doi.org/10.1038/nature11915} {\bibfield  {journal} {\bibinfo  {journal} {Nature}\ }\textbf {\bibinfo {volume} {495}},\ \bibinfo {pages} {210} (\bibinfo {year} {2013}{\natexlab{b}})}\BibitemShut {NoStop}%
\bibitem [{\citenamefont {Campos}\ \emph {et~al.}(1989)\citenamefont {Campos}, \citenamefont {Saleh},\ and\ \citenamefont {Teich}}]{PhysRevA.40.1371}%
  \BibitemOpen
  \bibfield  {author} {\bibinfo {author} {\bibfnamefont {R.~A.}\ \bibnamefont {Campos}}, \bibinfo {author} {\bibfnamefont {B.~E.~A.}\ \bibnamefont {Saleh}},\ and\ \bibinfo {author} {\bibfnamefont {M.~C.}\ \bibnamefont {Teich}},\ }\bibfield  {title} {\bibinfo {title} {Quantum-mechanical lossless beam splitter: Su(2) symmetry and photon statistics},\ }\href {https://doi.org/10.1103/PhysRevA.40.1371} {\bibfield  {journal} {\bibinfo  {journal} {Phys. Rev. A}\ }\textbf {\bibinfo {volume} {40}},\ \bibinfo {pages} {1371} (\bibinfo {year} {1989})}\BibitemShut {NoStop}%
\bibitem [{\citenamefont {Kim}\ \emph {et~al.}(2002)\citenamefont {Kim}, \citenamefont {Son}, \citenamefont {Bu\ifmmode~\check{z}\else \v{z}\fi{}ek},\ and\ \citenamefont {Knight}}]{PhysRevA.65.032323}%
  \BibitemOpen
  \bibfield  {author} {\bibinfo {author} {\bibfnamefont {M.~S.}\ \bibnamefont {Kim}}, \bibinfo {author} {\bibfnamefont {W.}~\bibnamefont {Son}}, \bibinfo {author} {\bibfnamefont {V.}~\bibnamefont {Bu\ifmmode~\check{z}\else \v{z}\fi{}ek}},\ and\ \bibinfo {author} {\bibfnamefont {P.~L.}\ \bibnamefont {Knight}},\ }\bibfield  {title} {\bibinfo {title} {Entanglement by a beam splitter: Nonclassicality as a prerequisite for entanglement},\ }\href {https://doi.org/10.1103/PhysRevA.65.032323} {\bibfield  {journal} {\bibinfo  {journal} {Phys. Rev. A}\ }\textbf {\bibinfo {volume} {65}},\ \bibinfo {pages} {032323} (\bibinfo {year} {2002})}\BibitemShut {NoStop}%
\bibitem [{Note1()}]{Note1}%
  \BibitemOpen
  \bibinfo {note} {The eigenstates for the undriven optomechanical system are exactly given by a two-mode Fock state displaced in the mechanical mode by a strength of $n\gamma _0/\Omega _m$, where $n$ is the photon number \cite {PhysRevA.91.013812}. As long as the microwave bath thermal population is very small or the optomechanical coupling is weak, our assumption that the system starts in a two mode thermal state is excellent.}\BibitemShut {Stop}%
\bibitem [{\citenamefont {Simon}(2000)}]{PhysRevLett.84.2726}%
  \BibitemOpen
  \bibfield  {author} {\bibinfo {author} {\bibfnamefont {R.}~\bibnamefont {Simon}},\ }\bibfield  {title} {\bibinfo {title} {Peres-horodecki separability criterion for continuous variable systems},\ }\href {https://doi.org/10.1103/PhysRevLett.84.2726} {\bibfield  {journal} {\bibinfo  {journal} {Phys. Rev. Lett.}\ }\textbf {\bibinfo {volume} {84}},\ \bibinfo {pages} {2726} (\bibinfo {year} {2000})}\BibitemShut {NoStop}%
\bibitem [{\citenamefont {Duan}\ \emph {et~al.}(2000)\citenamefont {Duan}, \citenamefont {Giedke}, \citenamefont {Cirac},\ and\ \citenamefont {Zoller}}]{PhysRevLett.84.2722}%
  \BibitemOpen
  \bibfield  {author} {\bibinfo {author} {\bibfnamefont {L.-M.}\ \bibnamefont {Duan}}, \bibinfo {author} {\bibfnamefont {G.}~\bibnamefont {Giedke}}, \bibinfo {author} {\bibfnamefont {J.~I.}\ \bibnamefont {Cirac}},\ and\ \bibinfo {author} {\bibfnamefont {P.}~\bibnamefont {Zoller}},\ }\bibfield  {title} {\bibinfo {title} {Inseparability criterion for continuous variable systems},\ }\href {https://doi.org/10.1103/PhysRevLett.84.2722} {\bibfield  {journal} {\bibinfo  {journal} {Phys. Rev. Lett.}\ }\textbf {\bibinfo {volume} {84}},\ \bibinfo {pages} {2722} (\bibinfo {year} {2000})}\BibitemShut {NoStop}%
\bibitem [{\citenamefont {Benito}\ \emph {et~al.}(2016)\citenamefont {Benito}, \citenamefont {S\'anchez Mu\~noz},\ and\ \citenamefont {Navarrete-Benlloch}}]{PhysRevA.93.023846}%
  \BibitemOpen
  \bibfield  {author} {\bibinfo {author} {\bibfnamefont {M.}~\bibnamefont {Benito}}, \bibinfo {author} {\bibfnamefont {C.}~\bibnamefont {S\'anchez Mu\~noz}},\ and\ \bibinfo {author} {\bibfnamefont {C.}~\bibnamefont {Navarrete-Benlloch}},\ }\bibfield  {title} {\bibinfo {title} {Degenerate parametric oscillation in quantum membrane optomechanics},\ }\href {https://doi.org/10.1103/PhysRevA.93.023846} {\bibfield  {journal} {\bibinfo  {journal} {Phys. Rev. A}\ }\textbf {\bibinfo {volume} {93}},\ \bibinfo {pages} {023846} (\bibinfo {year} {2016})}\BibitemShut {NoStop}%
\bibitem [{\citenamefont {Hughes}\ and\ \citenamefont {Dignam}(2024)}]{PhysRevA.110.063718}%
  \BibitemOpen
  \bibfield  {author} {\bibinfo {author} {\bibfnamefont {P.~R.~B.}\ \bibnamefont {Hughes}}\ and\ \bibinfo {author} {\bibfnamefont {M.~M.}\ \bibnamefont {Dignam}},\ }\bibfield  {title} {\bibinfo {title} {Analytic solution to the nonlinear generation of squeezed states in a thermal bath},\ }\href {https://doi.org/10.1103/PhysRevA.110.063718} {\bibfield  {journal} {\bibinfo  {journal} {Phys. Rev. A}\ }\textbf {\bibinfo {volume} {110}},\ \bibinfo {pages} {063718} (\bibinfo {year} {2024})}\BibitemShut {NoStop}%
\end{thebibliography}%

\end{document}